\documentclass[journal]{IEEEtran}
\usepackage{cite}

\ifCLASSINFOpdf
\else
\fi

\usepackage{color}
\usepackage{subfigure}
\usepackage[cmex10]{amsmath}
\usepackage[pdftex]{graphicx}
\usepackage{amsfonts}
\usepackage{bm}
\usepackage{makecell}
\usepackage{array}

\usepackage{graphicx}
\usepackage{color}
\usepackage{placeins}
\usepackage{float}
\usepackage{tabularx,colortbl}

\begin{document}
%
\title{Consensus and Sectioning-based ADMM with Norm-1 Regularization for Imaging with a Compressive Reflector Antenna}
%
%
%

\author{{Juan Heredia-Juesas$^{1,2}$, Ali Molaei$^{1}$, Luis Tirado$^{1}$, and Jos\'e \'A.
Mart\'inez-Lorenzo$^{1,2}$}
\thanks{$^{1}$Departments of Electrical \& Computer Engineering, Northeastern University, Boston, MA, USA. \tt\small jmartinez@coe.neu.edu}
\thanks{$^{2}$Departments of Mechanical \& Industrial Engineering, Northeastern University, Boston, MA, USA.}
}
\maketitle

\begin{abstract}
This paper presents three distributed techniques to find a sparse solution of the underdetermined linear problem $\textbf{g}=\textbf{Hu}$ with a norm-1 regularization, based on the Alternating Direction Method of Multipliers (ADMM). These techniques divide the matrix $\textbf{H}$ in submatrices by rows, columns, or both rows and columns, leading to the so-called \textit{consensus}-based ADMM, \textit{sectioning}-based ADMM, and \textit{consensus and sectioning}-based ADMM, respectively. These techniques are applied particularly for millimeter-wave imaging through the use of a Compressive Reflector Antenna (CRA). The CRA is a hardware designed to increase the sensing capacity of an imaging system and reduce the mutual information among measurements, allowing an effective imaging of sparse targets with the use of Compressive Sensing (CS) techniques. \textit{Consensus}-based ADMM has been proved to accelerate the imaging process and \textit{sectioning}-based ADMM has shown to highly reduce the amount of information to be exchange among the computational nodes. In this paper, the mathematical formulation and graphical interpretation of these two techniques, together with the \textit{consensus and sectioning}-based ADMM approach, are presented. The imaging quality, the imaging time, the convergence, and the communication efficiency among the nodes are analyzed and compared. The distributed capabitities of the ADMM-based approaches, together with the high sensing capacity of the CRA, allow the imaging of metallic targets in a 3D domain in quasi-real time with a reduced amount of information exchanged among the nodes.  
\end{abstract}


%
\IEEEpeerreviewmaketitle

\section{Introduction}
%
%
%
%
\IEEEPARstart{S}{everal} numerical techniques have been developed in the past decades for solving problems defined by a linear matrix equation \cite{bertsekas1989parallel}
\begin{equation}\label{linear_eq}
	\textbf{g}=\textbf{Hu},
\end{equation}
where $\textbf{g}\in\mathbb{C}^{m}$ and $\textbf{H}\in\mathbb{C}^{m\times n}$ are the known data, and $\textbf{u}\in\mathbb{C}^{n}$ is the unknown vector to be determined. These techniques can be classified in direct and iterative methods. Direct methods are capable of finding an exact solution of the equation (if existing) with a finite number of operations; but they may require an impractical amount of time. Iterative methods theoretically converge asymptotically to a solution with an infinite number of iterations; but an approximate solution, depending on the tolerance defined, can be achieved in a reduced amount of time. In both cases, the inversion of the matrix \textbf{H} or the matrix $\textbf{H}^*\textbf{H}$ is a problem that need to be addressed too, and also direct and iterative methods have been proposed to this end \cite{bertsekas1989parallel,greenspan1955methods,akaike1973block}. Despite the power enhancement of computational units, which reduce the operation times, the increase of data in recent years leads to a preference for iterative methods. Additionally, the presence of uncertainties or noise in the data is better addressed with the iterative methods since they find an approximate solution, that is, a solution within bounded limits. These uncertainties can be modeled by adding a noise vector $\textbf{w}\in\mathbb{C}^{m}$ to Eqn. \eqref{linear_eq} as follows:
\begin{equation}\label{linear_eq_with_noise}
\textbf{g}=\textbf{Hu}+\textbf{w}.
\end{equation}

Distributed techniques \cite{bertsekas1989parallel,boyd2011distributed,heredia2017norm,forero2010consensus,mota2012distributed,tsitsiklis1986distributed,olshevsky2009convergence,degroot1974reaching} allow to assign small pieces of information among several computational nodes for solving smaller problems in a parallel and fast fashion, exchanging the results among the nodes for obtaining a final solution. These distributed techniques may relief the computational load and speed up the convergence, but introduce the problem of communication among those computational nodes, which also has to be addressed \cite{fang2006communication,jakovetic2011cooperative,mehyar2005distributed,mota2013d,olfati2007consensus,schizas2008consensusI,schizas2008consensusII}. 

Regarding the properties of the unknown vector, of interest in the recent years are those underdetermined problems $(m\ll n)$ in which the solution sought is sparse; that is $\Vert\textbf{u}\Vert_0\ll n$, where $\Vert \cdot\Vert_0$ represents the number of non-zero elements of the vector. These type of problems are generally solved via the use of Compressive Sensing (CS) techniques by adding a norm-1 regularization, such as Bayesian Compressive Sampling (BCS) \cite{oliveri2011bayesian}, Fast Iterative Shrinkage-Thresholding Algorithm (FISTA) \cite{beck2009fast}, Nesterov's Algorithm (NESTA) \cite{Becker2011}, or the Alternating Direction Method of Multipliers (ADMM) \cite{boyd2011distributed,boyd2009convex,heredia2017norm,HerediaJuesas2015,mota2013d,erseghe2011fast}. 

This paper presents three iterative and distributive optimization techniques based on the ADMM to find a sparse solution of Eqn. \eqref{linear_eq_with_noise}, when the norm-1 regularization is applied. These techniques exploit the  distributed capabilities of the ADMM by dividing the matrix \textbf{H} in submatrices and solving the problem in several computational nodes. Dividing the matrix in submatrices by rows has shown in \cite{heredia2017norm} to reduce the time for finding a solution. In \cite{Heredia-JuesasUnderreview,HerediaJuesas2018fast} it has been proved that if the matrix is divided by columns, the amount of information to be shared among those computational units is highly reduced. This paper shows the mathematical formulation and graphical interpretation of the combination of the two previous techniques, with the aim of introducing more degrees of freedom for designing an appropriate optimization architecture.

Although these formulations are valid for any problem that could be represented in terms of the Eqn. \eqref{linear_eq_with_noise}, this paper shows their performance for a millimeter-wave imaging application through the use of a Compressive Reflector Antenna (CRA). A CRA is a hardware for increasing the sensing capacity of the imaging system, allowing a reduced number of measurement collection for performing imaging with the use of norm-1 regularized CS techniques \cite{molaei2017compressive,Martinez-Lorenzo2015,MartinezLorenzoInpress}. In this case $\textbf{H}\in {\mathbb{C}}^{N_m\times N_p}$ is called the sensing matrix, $\textbf{g}\in {\mathbb{C}}^{N_m}$ is the vector of measurements, and $\textbf{u}\in {\mathbb{C}}^{N_p}$ is the unknown vector of reflectivity, where $N_m$ represents the number of measurements collected and $N_p$ the number of pixels of the imaging domain.

This paper is organized as follows: Section \ref{ADMM_general_formulation} introduces the algorithm, properties, and conditions of the ADMM. Section \ref{ADMM_formulation} develops the mathematical formulation, the graphical interpretation, and the convergence process of the three presented methods for solving Eqn. \eqref{linear_eq_with_noise}:
\begin{itemize}
	\item \textit{Consensus-based ADMM}. Dividing the sensing matrix in submatrices by rows.
	\item \textit{Sectioning-based ADMM}. Dividing the sensing matrix in submatrices by columns.
	\item \textit{Consensus and sectioning-based ADMM}. Dividing the sensing matrix in submatrices by rows and columns.
\end{itemize}
Section \ref{Communications} studies the communications among the computational nodes, comparing the amount of information exchanged by one single node at one iteration for the three different techniques. Section \ref{CRA} briefly introduces the description and operation of the CRA. The particular configuration and numerical results are shown in section \ref{Results}, where the imaging quality, imaging time, convergence, and communication efficiency among the computational nodes are compared and discussed for the three proposed techniques. The paper concludes in section \ref{conclusion}.

\section{General formulation of the ADMM}\label{ADMM_general_formulation}
The ADMM is an optimization algorithm for convex functions that takes advantage of both the dual ascend decomposability, spliting the objective function into simpler objectives, and the convergence properties of the method of multipliers, which relaxes the conditions of the objective function. The general representation of the ADMM takes the following optimization form \cite{boyd2011distributed, boyd2009convex}:
\begin{equation} \label{ADMM_general}
\left.
\begin{array}{c}
\mbox{minimize $f_1(\textbf{u})+f_2(\textbf{v})$} \\
\mbox{$\;\;\;$s.t.$\;\;\;\;\;\;\textbf{Pu}+\textbf{Qv}=\textbf{c}$},%
\end{array}%
\right.
\end{equation}%
where the known matrices $\textbf{P}\in{\mathbb{C}}^{p \times n}$ and $\textbf{Q}\in{\mathbb{C}}^{p \times q}$, and vector $\textbf{c}\in{\mathbb{C}}^{p}$ determine the constraint over the unknown variable vectors $\textbf{u}\in{\mathbb{C}}^{n}$ and $\textbf{v}\in{\mathbb{C}}^{q}$. The convex functions $f_1$ and $f_2$ have to be extended real value functions, that is 
\begin{subequations}
	\begin{equation}
		f_1:\mathbb{C}^n\rightarrow\mathbb{R}\cup \{+\infty\},
	\end{equation}
	\begin{equation}
		f_2:\mathbb{C}^q\rightarrow\mathbb{R}\cup \{+\infty\},
	\end{equation}
\end{subequations}
and they have to be closed and proper, that is, their effective domain (non-infinity values) has to be non-empty and they never reach $-\infty$, mathematically:
\begin{subequations}
	\begin{equation}
		\exists \textbf{u}\in Dom \{f\}\; | \; f(\textbf{u})<+\infty, \;\; \text{and}
	\end{equation}
	\begin{equation}
		f(\textbf{u})>-\infty, \; \forall \textbf{u}\in Dom \{f\},
	\end{equation}
\end{subequations}

The optimal value of \eqref{ADMM_general} may be denoted by $\textbf{t}^{\star}$ as
\begin{equation}
	\textbf{t}^{\star}=\text{inf}\big\{f_1(\textbf{u})+f_2(\textbf{v})\;|\;\textbf{Pu}+\textbf{Qv}=\textbf{c}\big\}.
\end{equation}
Taking advantage of the method of multipliers \cite{bertsekas2014constrained}, the augmented Lagrangian form of this problem is defined as follows:
\begin{eqnarray}
	L_{\rho}\left( \textbf{u},\textbf{v},\textbf{d}\right)= f_1(\textbf{u})+f_2(\textbf{v}) + \notag \\
	+\textbf{d}^T \left(\textbf{Pu}+\textbf{Qv}-\textbf{c}\right)+\frac{\rho}{2}\Vert \textbf{Pu}+\textbf{Qv}-\textbf{c}\Vert_2^2, 
\end{eqnarray}
where $\textbf{d}\in{\mathbb{C}}^{p}$ is the Lagrangian multiplier or dual variable, and $\rho>0$ is the augmented parameter. A more convenient expression of the augmented Lagrangian can be achieved by the following simple algebraic transformation:
\begin{equation}
	\textbf{d}^T\textbf{r}+\frac{\rho}{2}\Vert \textbf{r} \Vert_2^2=\frac{\rho}{2}\Vert \textbf{r}+\textbf{s} \Vert_2^2-\frac{\rho}{2}\Vert \textbf{s} \Vert_2^2,
\end{equation}
for $\textbf{r}=\textbf{Pu}+\textbf{Qv}-\textbf{c}$, and $\textbf{s}=1/\rho\;\textbf{d}$ being the scaled dual variable.
Based on this, the general iterative algorithm of the ADMM is described as 
\begin{subequations}
	\begin{align}
	\textbf{u}^{(k+1)} &:=\underset{\textbf{u}}{\text{argmin }} L_{\rho}\left(\textbf{u},\textbf{v}^{(k)},\textbf{s}^{(k)}\right),\\
	\textbf{v}^{(k+1)} &:=\underset{\textbf{v}}{\text{argmin }} L_{\rho}\left(\textbf{u}^{(k+1)},\textbf{v},\textbf{s}^{(k)}\right),\\
	\textbf{s}^{(k+1)} &:=\textbf{s}^{(k)}+\left(\textbf{Pu}^{(k+1)}+\textbf{Qv}^{(k+1)}-\textbf{c}\right).
	\end{align}%
\end{subequations}
The fact that $f_1$ and $f_2$ are defined over different variables allows the optimization of \textbf{u} and \textbf{v} in an \textit{alternating direction} fashion. 

Two metrics are defined for evaluating the convergence of the ADMM algorithm. The \textit{primal residual}, which measures the residual of the constraint; and the \textit{dual residual}, which measures the residual of the dual variable optimization between two consecutive iterations, are defined, respectively at iteration $k$, as follows \cite{boyd2011distributed}:
\begin{subequations}
	\begin{align}
		\textbf{r}_{\textbf{p}}^{(k)}&=\textbf{Pu}^{(k)}+\textbf{Qv}^{(k)}-\textbf{c}, \\
		\textbf{r}_{\textbf{d}}^{(k)}&=\rho\textbf{P}^T\textbf{Q}\left(\textbf{v}^{(k)}-\textbf{v}^{(k-1)}\right)
	\end{align}%
\end{subequations}

\section{ADMM distributed solving methods}\label{ADMM_formulation}
ADMM is a convenient method when applying CS for solving Eqn. \eqref{linear_eq_with_noise}. Under the assumption that the sensing matrix \textbf{H} satisfies the Restricted Isometry Property (RIP) \cite{candes2008restricted, Obermeier2016model}, and that the unknown vector \textbf{u} is sparse---that is, the number of non-zero elements $N_{nz}$ is much smaller than the total number of elements, $N_{nz}\ll n$---Eqn. \eqref{linear_eq_with_noise} can be solved by minimizing the sum of the convex function $f_1(\textbf{u})=\frac{1}{2}\left\Vert \textbf{Hu}-\textbf{g}\right\Vert _{2}^{2}$ and the norm-1 regularization $f_2(\textbf{v})=\lambda \left\Vert \textbf{v}\right\Vert _{1}$.  The particular ADMM formulation for solving Eqn. \eqref{linear_eq_with_noise} takes the \textit{lasso} form:
\begin{equation}
	\left.
	\begin{array}{cc}
	\mbox{minimize} & \frac{1}{2}\left\Vert \textbf{Hu}-\textbf{g}\right\Vert _{2}^{2}+\lambda \left\Vert \textbf{v}\right\Vert _{1} \\
	\mbox{s.t.} & \textbf{u}-\textbf{v}=\textbf{0}.%
	\end{array}
	\right.
	\label{original}
\end{equation}
The constrain---defined with $\bf{P=I}$, $\bf{Q=-I}$, and $\bf{c=0}$---enforces that the variables $\textbf{u}$ and $\textbf{v}$ are equal.

Since the dimensions of the sensing matrix \textbf{H} could be very large---having many pixels in the imaging domain and/or many collected measurements---, a direct resolution of the problem \eqref{original} is not usually efficient. Some techniques have been proposed for solving this problem in a distributed fashion using the ADMM, such as \cite{heredia2017norm} or \cite{Heredia-JuesasUnderreview}, for solving fast imaging problems
; or \cite{mota2012distributed, mota2013d}, 
for solving a communications problem in the dual space. In this paper, three different methods, focused on solving imaging problems in the primal space, are presented. The aim is to find a sparse solution of Eqn. \eqref{linear_eq_with_noise}, while reducing the amount of information exchanged among the nodes, and the computational complexity and time. 

\subsection{Consensus-based ADMM: Row-wise division}\label{ConsensusADMM}
As presented in \cite{heredia2017norm}, problem \eqref{original} can be solved in a distributed fashion, by splitting the original matrix \textbf{H} into $M$ submatrices $\textbf{H}_i\in\mathbb{C}^{\frac{N_m}{M}\times N_p}$ in a row division, and the vector of measurements \textbf{g} into $M$ subvectors $\textbf{g}_i\in\mathbb{C}^{\frac{N_m}{M}}$, as shown in Fig. \ref{ByRows}a. Then, $M$ different underdetermined problems $\textbf{H}_i\textbf{u}=\textbf{g}_i$, for $i=1,\dots ,M$, need to be solved. In particular, the summation of all of them may be optimized together with the norm-1 regularization as follows:
\begin{equation}
	\left.
	\begin{array}{cc}
	\mbox{minimize } & \frac{1}{2}\sum\limits_{i=1}^{M}\left\Vert
	\textbf{H}_{i}\textbf{u}-\textbf{g}_{i}\right\Vert _{2}^{2}+\lambda\left\Vert \textbf{v}\right\Vert _{1} \\
	\mbox{s.t.} & \textbf{u}=\textbf{v}.%
	\end{array}%
	\right.
	\label{by_rows_no_consensus}
\end{equation}
In order to make the optimizations independent, $M$ replicas of the unknown variable \textbf{u} may be defined as $\textbf{u}^i$ for $i=1,\dots,M$, turning the expression \eqref{by_rows_no_consensus} into 
\begin{equation}
	\left.
	\begin{array}{cc}
	\mbox{minimize } & \frac{1}{2}\sum\limits_{i=1}^{M}\left\Vert
	\textbf{H}_{i}\textbf{u}^i-\textbf{g}_{i}\right\Vert _{2}^{2}+\lambda\left\Vert \textbf{v}\right\Vert _{1} \\
	\mbox{s.t.} & \textbf{u}^i=\textbf{v},\;\;\forall i\in\{1,...,M\}.%
	\end{array}%
	\right.
	\label{by_rows_consensus}
\end{equation}
The augmented Lagrangian function for this problem is as follows:
\begin{eqnarray}
	L_{\rho}\left( \textbf{u}^1,\dots,\textbf{u}^M,\textbf{v},\textbf{s}^1,\dots,\textbf{s}^M\right)=  \notag \\  =\frac{1}{2}\sum\limits_{i=1}^{M}\left\Vert
	\textbf{H}_{i}\textbf{u}^i-\textbf{g}_{i}\right\Vert _{2}^{2}+\lambda\left\Vert \textbf{v}\right\Vert _{1}+  \\
	+\frac{\rho }{2}\sum\limits_{i=1}^{M}\left\Vert \textbf{u}^i-\textbf{v}+\textbf{s}^i\right\Vert _{2}^{2}-\frac{\rho }{2}\sum%
	\limits_{i=1}^{M}\left\Vert \textbf{s}^i\right\Vert _{2}^{2}, \notag
\end{eqnarray}
where a dual variable $\textbf{s}^i$ is introduced for each of the $M$ constraints. The augmented parameter $\rho $ enforces the convexity of the Lagrangian function. By iterating the following scheme, an optimal solution may be found: 
\begin{figure}[htp]
	\centering
	\vspace{0pt}
	\includegraphics[scale=.48, trim = 6mm 40mm 0mm 10mm, clip]{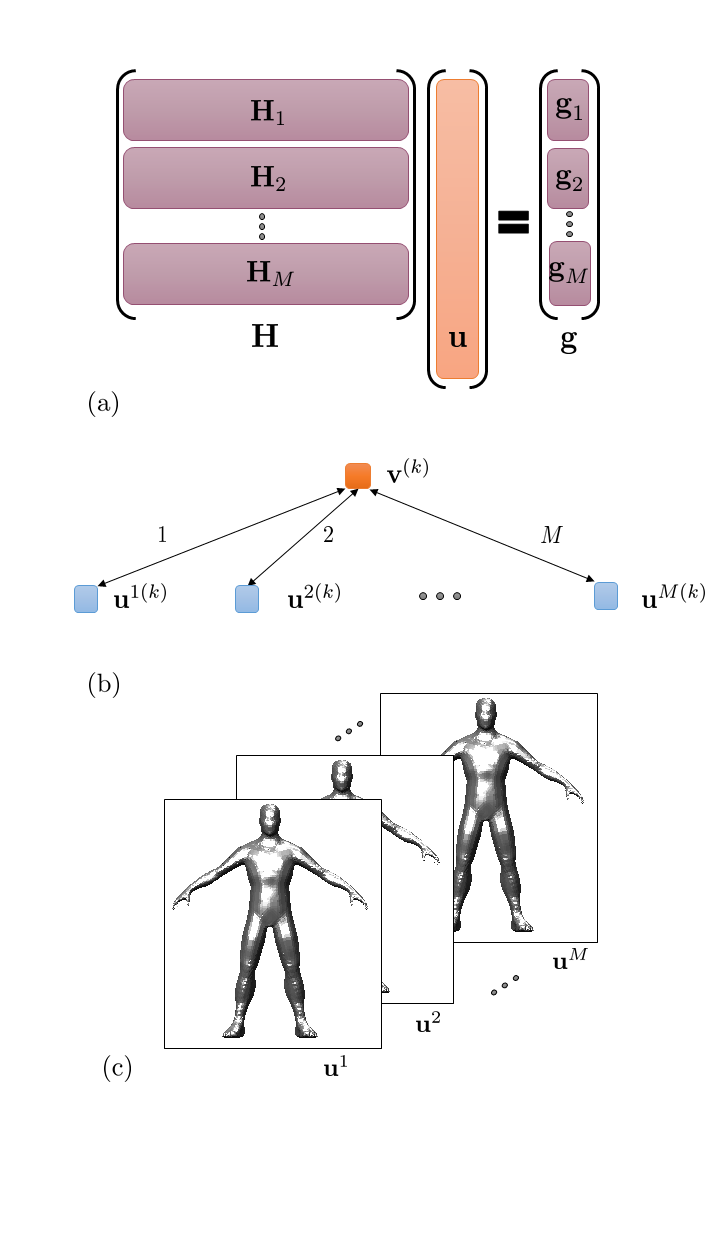}
	\caption{(a) Division of the matrix equation system by rows. (b) Architecture of the consensus-based ADMM: a central node collects the updates of $M$ sub-nodes, computes the soft-thresholding operator of the mean of them, and then distributes the solution again to the sub-nodes. (c) Graphical interpretation of the row-wise division: $M$ independent images are optimized with few data allocated to each node. The final imaging is an average-like of all of them.}
	\vspace{0pt}
	\label{ByRows}
\end{figure}
\begin{subequations}
\begin{align}
	\label{x_solution}
	\textbf{u}^{i,(k+1)} &=\left( \textbf{H}_{i}^{\ast }\textbf{H}_{i}+\rho \textbf{I}_{N_p}\right) ^{-1}\left(\textbf{H}_{i}^{\ast }\textbf{g}_{i}+\rho \left(\textbf{v}^{(k)}-\textbf{s}^{i(k)}\right) \right) , \\
	\textbf{v}^{(k+1)} &=\mathbf{S}_{\frac{\lambda}{M\rho}}\left( \bar{\textbf{u}}^{(k+1)}+\bar{\textbf{s}}^{(k)}\right) , \label{v_variable}\\
	\textbf{s}^{i(k+1)} &=\textbf{s}^{i(k)}+\textbf{u}^{i(k+1)}-\textbf{v}^{(k+1)} ,
\end{align}%
\end{subequations}
where $\bar{\textbf{u}}$ and $\bar{\textbf{s}}$ represent the mean of $\textbf{u}^i$ and $\textbf{s}^i$, respectively, for all values of $i$; $\textbf{I}_{N_p}$ indicate the identity matrix of size $N_p$; and $\mathbf{S}_{\kappa }\left( a\right) $ is the element-wise soft thresholding
operator \cite{bredies2008linear}:
\begin{align}
\label{softhresholding}
\mathbf{S}_{\kappa }
(a) = \begin{cases}
a - \kappa \operatorname{sign}(a), & |a| > \kappa \\
0 & |a| \le \kappa.
\end{cases}
\end{align} 
The \textit{matrix inversion lemma} \cite{woodbury1950inverting} may be applied for the computation of the term $\left( \textbf{H}_{i}^{\ast }\textbf{H}_{i}+\rho \textbf{I}_{N_p}\right) ^{-1}$, as shown in Eqn. \eqref{inversion_lemma}. Therefore, just inverting $M$ matrices of reduced size $\frac{N_m}{M}\times \frac{N_m}{M}$, instead of $M$ large matrices of size $N_p\times N_p$, is required, highly accelerating the algorithm.
\begin{equation}
\label{inversion_lemma}
	\left( \textbf{H}_{i}^{\ast }\textbf{H}_{i}+\rho \textbf{I}_{N_p}\right) ^{-1}=\frac{\textbf{I}_{N_p}}{\rho }-
	\frac{\textbf{H}_{i}^{\ast}}{\rho^{2}}\left(\textbf{I}_{\frac{N_m}{M}}+\frac{\textbf{H}_{i}\textbf{H}_{i}^{\ast}}
	{\rho}\right) ^{-1}\textbf{H}_{i},
\end{equation}

In terms of convergence, the primal and dual residuals are computed, respectively, as follows:
\begin{subequations}
	\begin{equation}
		\textbf{r}_{\textbf{p}}^{(k)}=\left(\textbf{u}^{1,(k)}-\textbf{v}^{(k)},\dots,\textbf{u}^{M,(k)}-\textbf{v}^{(k)}\right),
	\end{equation}
	\begin{equation}
		\textbf{r}_{\textbf{d}}^{(k)}=-\rho\left(\textbf{v}^{(k)}-\textbf{v}^{(k-1)},\dots,\textbf{v}^{(k)}-\textbf{v}^{(k-1)}\right),
	\end{equation}
\end{subequations}
and their squared norms are
\begin{subequations}
	\begin{equation}\label{primal_consensus}
		\Vert\textbf{r}_{\textbf{p}}^{(k)}\Vert_2^2=\sum_{i=1}^M\Vert \textbf{u}^{i,(k)}-\textbf{v}^{(k)}\Vert_2^2,
	\end{equation}	
	\begin{equation}
		\Vert\textbf{r}_{\textbf{d}}^{(k)}\Vert_2^2=\rho^2M\Vert \textbf{v}^{(k)}-\textbf{v}^{(k-1)}\Vert_2^2,
	\end{equation}	
\end{subequations}
noticing that Eqn. \eqref{primal_consensus} can be interpreted as a measure of the \textit{lack of consensus}.

It can be noticed in expressions \eqref{by_rows_consensus} and \eqref{v_variable} that the variable \textbf{v} acts as a \textit{consensus}, forcing that all variables $\textbf{u}^i$ converge to the same solution. The architecture of this algorithm can be interpreted as a hierarchical structure, having a central node that collects all individual solution for each sub-node, performs the soft-thresholding averaging, and then broadcasts the global solution to each sub-node, as represented in Fig. \ref{ByRows}b. The purpose of this technique is to perform $M$ independent images with few amount of data allocated to each node, and then create the final imaging as an average-like of the intermediate results, in the manner that Fig. \ref{ByRows}c shows.

As shown in \cite{heredia2017norm}, this technique highly reduces the computational cost producing real-time imaging; however, it has the problem of sharing the global solution $\textbf{v}^{(k)}$ from the central node to each sub-node, and the whole individual solution $\textbf{u}^{i,(k+1)}$ from each sub-node to the central node, for each iteration. These vectors are of the size of the total number of pixels in the imaging domain and may be very large, producing a slow communication among the computational nodes. 

\subsection{Sectioning-based ADMM: Column-wise division}
\begin{figure}[htp]
	\centering
	\vspace{0pt}
	\includegraphics[scale=.50, trim = 6mm 60mm 0mm 5mm, clip]{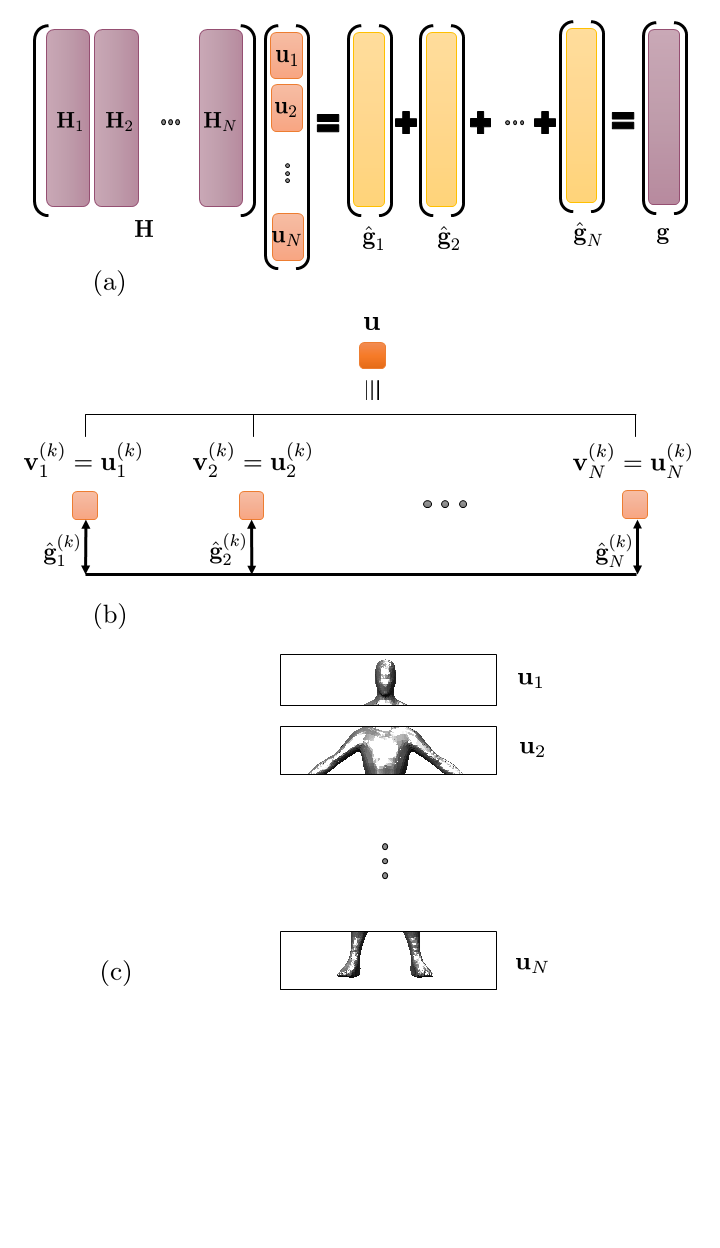}
	\caption{(a) Division of the matrix equation system by columns. The measurements vector is decomposed in $N$ estimated vectors. (b) Architecture of the sectioning-based ADMM: the problem is split into $N$ nodes that optimize a part of the imaging. For each iteration, they share the small estimated data vector with the remaining nodes. (c) Graphical interpretation of the ADMM column-wise division: the image is sectioned into $N$ regions. The final imaging is the concatenation of all of them.}
	\vspace{0pt}
	\label{ByColumns}
\end{figure}

A different approach for finding a solution of problem \eqref{original} is by dividing the original matrix \textbf{H} into $N$ submatrices $\textbf{H}_j\in\mathbb{C}^{{N_m}\times \frac{N_p}{N}}$ in a column basis and, accordingly, the vector of unknowns \textbf{u} into $N$ subvectors $\textbf{u}_j\in\mathbb{C}^\frac{N_p}{N}$, as done in \cite{Heredia-JuesasUnderreview}. This segmentation makes the problem to be solved in the following form: $\sum_{j=1}^N\textbf{H}_j\textbf{u}_j=\sum_{j=1}^N\hat{\textbf{g}}_j=\textbf{g}$, which requires the introduction of the so-called \textit{estimated data} vectors $\hat{\textbf{g}}_j$,
as represented in Fig. \ref{ByColumns}a. The problem is optimized, together with the norm-1 regularization, as follows:
\begin{equation}
	\left.
	\begin{array}{cc}
	\mbox{minimize } & \frac{1}{2}\left\Vert\sum\limits_{j=1}^{N}
	\textbf{H}_{j}\textbf{u}_j-\textbf{g}\right\Vert _{2}^{2}+\lambda\sum\limits_{j=1}^N\left\Vert \textbf{v}_j\right\Vert _{1} \\
	\mbox{s.t.} & \textbf{u}_j=\textbf{v}_j,\;\;\forall j\in\{1,...,N\}.%
	\end{array}%
	\right.
	\label{by_columns_no_consensus}
\end{equation}
The augmented Lagrangian for this problem is defined over $3N$ variables as in the following expression:
\begin{eqnarray}
	L_{\rho}\left( \textbf{u}_1,\dots,\textbf{u}_N,\textbf{v}_1,\dots,\textbf{v}_N,\textbf{s}_1,\dots,\textbf{s}_N\right) = \notag \\ =\frac{1}{2}\left\Vert\sum\limits_{j=1}^{N}
	\textbf{H}_{j}\textbf{u}_j-\textbf{g}\right\Vert _{2}^{2}+\lambda\sum\limits_{j=1}^N\left\Vert \textbf{v}_j\right\Vert _{1}+  \\
	+\frac{\rho }{2}\sum\limits_{j=1}^{N}\left\Vert \textbf{u}_j-\textbf{v}_j+\textbf{s}_j\right\Vert _{2}^{2}-\frac{\rho }{2}\sum%
	\limits_{j=1}^{N}\left\Vert \textbf{s}_j\right\Vert _{2}^{2}, \notag
\end{eqnarray}
where, again, $\textbf{s}_j$ is the dual variable introduced for each constraint $j$, and $\rho$ is the augmented parameter.
This problem can be solved by the following iterative scheme:
\begin{subequations}
	\begin{align}
	\label{x_solution_columns}
	\textbf{u}_j^{(k+1)} &=\left( \textbf{H}_{j}^{\ast }\textbf{H}_{j}+\rho \textbf{I}_{\frac{N_p}{N}}\right) ^{-1}\left(\textbf{H}_{j}^{\ast }\textbf{g}_{j}^{(k)}+\rho \left(\textbf{v}_j^{(k)}-\textbf{s}_j^{(k)}\right) \right) , \\
	\textbf{v}_j^{(k+1)} &=\mathbf{S}_{\frac{\lambda}{\rho}}\left( \textbf{u}_j^{(k+1)}+\textbf{s}_j^{(k)}\right) , \label{v_variable_columns}\\
	\textbf{s}_j^{(k+1)} &=\textbf{s}_j^{(k)}+\textbf{u}_j^{(k+1)}-\textbf{v}_j^{(k+1)} ,
	\end{align}
\end{subequations}%
where $\textbf{g}_j^{(k)}$, required for computing Eqn. \eqref{x_solution_columns}, is obtained as
\begin{equation}
	\label{g_columns}
	\textbf{g}_j^{(k)}=\textbf{g}-\sum_{\substack{q=1 \\ q\neq j}}^N\textbf{H}_q\textbf{u}_q^{(k)} = \textbf{g}-\sum_{\substack{q=1 \\ q\neq j}}^N \hat{\textbf{g}}_q^{(k)},
\end{equation}
and it corresponds with the fraction of data determined for the update of the segment $j$ of the vector $\textbf{u}$, taking into account the \textit{estimated data} computed from the remaining nodes. $\mathbf{S}_{\kappa }\left( a\right) $ is the soft thresholding operator as defined in Eqn. \eqref{softhresholding}. In case of $N_m<\frac{N_p}{N}$, the \textit{matrix inversion lemma} can be applied to the term $\left(\textbf{H}_{j}^{\ast }\textbf{H}_{j}+\rho \textbf{I}_{\frac{N_p}{N}}\right)^{-1}$ as follows: 
\begin{equation}
	\label{inversion_lemma2}
	\left( \textbf{H}_{j}^{\ast }\textbf{H}_{j}+\rho \textbf{I}_\frac{N_p}{N}\right) ^{-1}=\frac{\textbf{I}_\frac{N_p}{N}}{\rho }-\frac{%
	\textbf{H}_{j}^{\ast }}{\rho ^{2}}\left( \textbf{I}_{N_m}+\frac{\textbf{H}_{j}\textbf{H}_{j}^{\ast }}{\rho }%
	\right) ^{-1}\textbf{H}_{j}.
\end{equation}
In this case, only $N$ matrices of sizes $N_m\times N_m$ need to be inverted. However, if $N_m>\frac{N_p}{N}$, the original inversion is computationally more efficient. 

In terms of convergence, the primal and dual residual vectors for this technique are computed, respectively, as follows:
\begin{subequations}
	\begin{equation}
	\textbf{r}_{\textbf{p}}^{(k)}=\left(\textbf{u}_1^{(k)}-\textbf{v}_1^{(k)},\dots,\textbf{u}_N^{(k)}-\textbf{v}_N^{(k)}\right),
	\end{equation}
	\begin{equation}
	\textbf{r}_{\textbf{d}}^{(k)}=-\rho\left(\textbf{v}_1^{(k)}-\textbf{v}_1^{(k-1)},\dots,\textbf{v}_N^{(k)}-\textbf{v}_N^{(k-1)}\right);
	\end{equation}
\end{subequations}
and their squared norms are
\begin{subequations}
	\begin{equation}
	\Vert\textbf{r}_{\textbf{p}}^{(k)}\Vert_2^2=\sum_{j=1}^N\Vert \textbf{u}_j^{(k)}-\textbf{v}_j^{(k)}\Vert_2^2,
	\end{equation}	
	\begin{equation}
	\Vert\textbf{r}_{\textbf{d}}^{(k)}\Vert_2^2=\rho^2\sum_{j=1}^N\Vert \textbf{v}_j^{(k)}-\textbf{v}_j^{(k-1)}\Vert_2^2.
	\end{equation}	
\end{subequations}

It is deducted from the analysis of Eqns. \eqref{x_solution_columns} and \eqref{g_columns} that, for performing the $\textbf{u}^{(k+1)}_j$ optimizations, each computational node $j$ needs the submatrix $\textbf{H}_j$, the whole vector \textbf{g}, and the \textit{estimated data} coming from the remaining nodes $\hat{\textbf{g}}_q^{(k)}$, for $q\neq j$. Therefore, this problem can be interpreted as an $N$ fully-connected net of nodes that individually optimize each fragment $\textbf{u}^{(k+1)}_j$, introducing thereupon its update to the net in the format of the \textit{estimated data} $\hat{\textbf{g}}^{(k+1)}_j=\textbf{H}_j\textbf{u}^{(k+1)}_j\in\mathbb{C}^{N_m}$, creating a non-hierarchical architecture, as the one represented in Fig. \ref{ByColumns}b. This approach can be illustrated as a \textit{sectioning} of the imaging domain due to splitting the unknown vector \textbf{u} into $N$ subvectors, corresponding each $\textbf{u}_j$ to a specific region of the image, as it is schematized in Fig. \ref{ByColumns}c. These regions may be predetermined by the user by an appropriate division of the unknown vector \textbf{u} and, consequently, the matrix \textbf{H} would be divided accordingly. The final imaging solution is accomplished by connecting the $N$ optimizations $\textbf{u}=[\textbf{u}_1;\dots; \textbf{u}_N]$.

This technique takes advantage of this image sectioning, since the communication among the nodes requires sharing only small vectors $\hat{\textbf{g}}^{(k)}_q\in\mathbb{C}^{N_m}$, for each iteration $k$. However, it lacks the acceleration achieved in the row-wise division due to two main reasons: (i) for small values of $N$, the inversion of the matrices in Eqn. \eqref{inversion_lemma2} might be expensive, and (ii) for large values of $N$, the known vector of measurements \textbf{g} is highly scattered into the $N$ estimations $\hat{\textbf{g}}_j$, 
causing slow computation at each iteration because of the matrix-vector product.

\subsection{Consensus and sectioning-based ADMM. Row and column-wise division}
\begin{figure}[htp]
	\centering
	\vspace{-5pt}
	\includegraphics[scale=.44, trim = 2mm 260mm 0mm 0mm, clip]{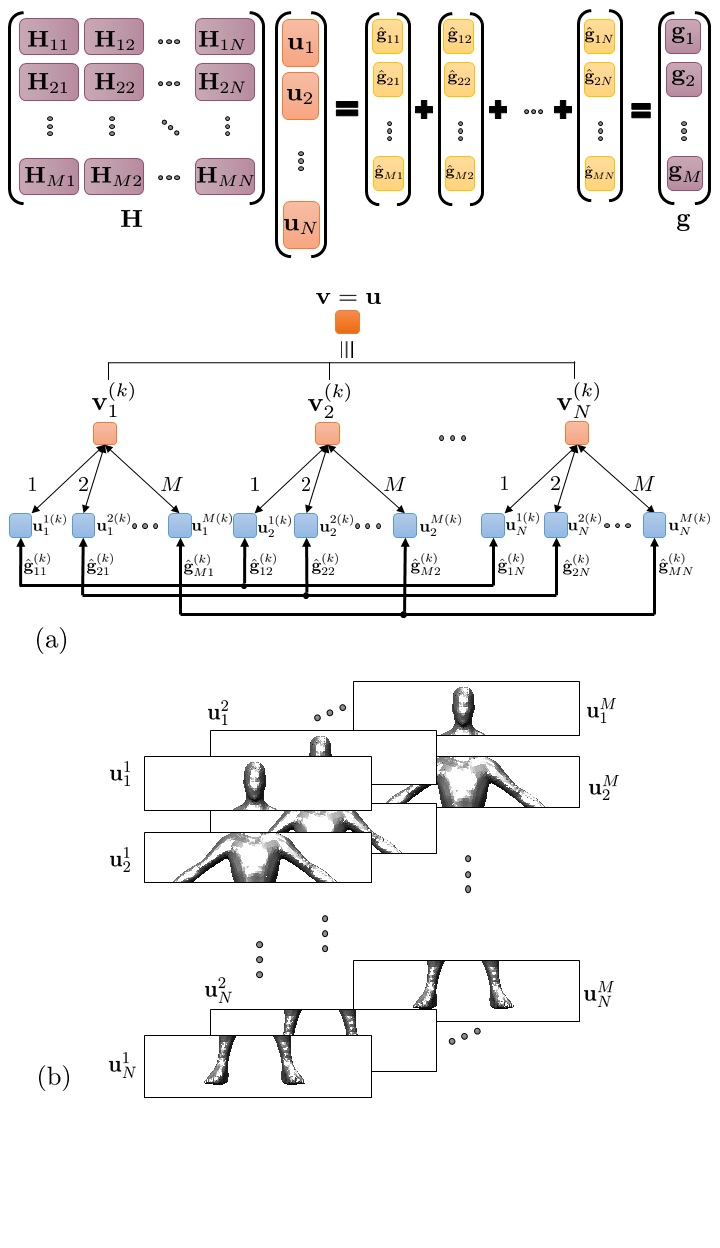}
	\caption{Division of the matrix equation system by rows and columns. The measurements vector is divided into $M$ subvectors and each of them is decomposed into $N$ estimated vectors.}
	\vspace{-5pt}
	\label{ByRows_Columns}
\end{figure}

A combination of the two previous approaches may be performed when dividing the matrix \textbf{H} into $M\cdot N$ submatrices $\textbf{H}_{ij}\in\mathbb{C}^{\frac{N_m}{M}\times \frac{N_p}{N}}$, the vector of measurements \textbf{g} into $M$ subvectors $\textbf{g}_i\in\mathbb{C}^{\frac{N_m}{M}}$, and the unknown vector \textbf{u} into $N$ subvectors $\textbf{u}_j\in\mathbb{C}^\frac{N_p}{N}$, as shown in Fig. \ref{ByRows_Columns}. Now, $M$ underdetermined problems $\sum_{j=1}^N\textbf{H}_{ij}\textbf{u}_j=\sum_{j=1}^N\hat{\textbf{g}}_{ij}=\textbf{g}_i$, for $i=1,\dots,M$, need to be solved. Applying the same technique as in the division by rows, that is, minimizing the summation of all of them and creating $M$ replicas of each segment $j$ of the unknown vector \textbf{u}, namely $\textbf{u}_j^i$, the problem may be optimized, together with the norm-1 regularization, as follows:
\begin{equation}
\left.
\begin{array}{cc}
\mbox{minimize } & \frac{1}{2}\sum\limits_{i=1}^M\left\Vert\sum\limits_{j=1}^{N}
\textbf{H}_{ij}\textbf{u}^i_j-\textbf{g}_i\right\Vert _{2}^{2}+\lambda\sum\limits_{j=1}^N\left\Vert \textbf{v}_j\right\Vert _{1} \\
\mbox{s.t.} & \textbf{u}_j^i=\textbf{v}_j,\;\;\forall i\in\{1,...,M\},\;\;\forall j\in\{1,...,N\}.%
\end{array}%
\right.
\label{by_rows_and_columns_consensus}
\end{equation}
Notice that this problem has $M\cdot N$ equality constraints. 

The augmented Lagrangian function for this problem, with $\left(2M+1\right)N$ variables, is expressed in the next equation:
\begin{eqnarray}
	L_{\rho}\left( \textbf{u}^1_1,\dots,\textbf{u}^M_N,\textbf{v}_1,\dots,\textbf{v}_N,\textbf{s}^1_1,\dots,\textbf{s}^M_N\right) = \notag \\ =\frac{1}{2}\sum\limits_{i=1}^M\left\Vert\sum\limits_{j=1}^{N}
	\textbf{H}_{ij}\textbf{u}^i_j-\textbf{g}_i\right\Vert _{2}^{2}+\lambda\sum\limits_{j=1}^N\left\Vert \textbf{v}_j\right\Vert _{1}+   \\
	+\frac{\rho }{2}\sum\limits_{i=1}^{M}\sum\limits_{j=1}^{N} \left\Vert \textbf{u}^i_j-\textbf{v}_j+\textbf{s}^i_j\right\Vert _{2}^{2}-\frac{\rho }{2}\sum\limits_{i=1}^{M}\sum%
	\limits_{j=1}^{N}\left\Vert \textbf{s}^i_j\right\Vert _{2}^{2}, \notag
\end{eqnarray}
where $\textbf{s}^i_j$ is the dual variable for the constraint with indices $i$ and $j$, and $\rho $ is, as in previous cases, the augmented parameter.
This problem can be solved by the following iterative scheme:
\begin{subequations}
\begin{align}
\label{x_solution_rows_columns}
\textbf{u}_j^{i,(k+1)} &=\left( \textbf{H}_{ij}^{\ast }\textbf{H}_{ij}+\rho \textbf{I}_{\frac{N_p}{N}}\right) ^{-1}\left(\textbf{H}_{ij}^{\ast }\textbf{g}_{ij}^{(k)}+\rho \left(\textbf{v}_j^{(k)}-\textbf{s}_j^{i,(k)}\right) \right) , \\
\textbf{v}_j^{(k+1)} &=\mathbf{S}_{\frac{\lambda}{M\rho}}\left( \bar{\textbf{u}}_j^{(k+1)}+\bar{\textbf{s}}_j^{(k)}\right) , \label{v_variable_rows_columns}\\
\textbf{s}_j^{i,(k+1)} &=\textbf{s}_j^{i,(k)}+\textbf{u}_j^{i,(k+1)}-\textbf{v}_j^{(k+1)} , \label{s_variable_rows_columns}
\end{align}%
\end{subequations}
where 
\begin{equation}
\label{g_rows_columns}
\textbf{g}_{ij}^{(k)}=\textbf{g}_i-\sum_{\substack{q=1 \\ q\neq j}}^N\textbf{H}_{iq}\textbf{u}_q^{i,(k)} = \textbf{g}_i-\sum_{\substack{q=1 \\ q\neq j}}^N \hat{\textbf{g}}_{iq}^{(k)},
\end{equation}
corresponds with the fraction of data determined for the update of the $i-th$ replica of the segment $j$ of the vector $\textbf{u}$, which takes into account the \textit{estimated data} computed for the remaining nodes of the same replica $i$.
$\mathbf{S}_{\kappa }\left( a\right) $ is the soft thresholding operator as defined in Eqn. \eqref{softhresholding}, and $\bar{\textbf{u}}_j$ and $\bar{\textbf{s}}_j$ are the mean of $\textbf{u}^i_j$ and $\textbf{s}^i_j$, respectively, for all replicas $i$ of a given segment $j$. 
If $\frac{N_m}{M}<\frac{N_p}{N}$, the \textit{matrix inversion lemma} should be applied for the inversion of the term $\left( \textbf{H}_{ij}^{\ast }\textbf{H}_{ij}+\rho \textbf{I}_{\frac{N_p}{N}}\right) ^{-1}$, as indicated in the Eqn. \eqref{inversion_lemma3}:
\begin{equation}
\label{inversion_lemma3}
\left( \textbf{H}_{ij}^{\ast }\textbf{H}_{ij}+\rho \textbf{I}_\frac{N_p}{N}\right) ^{-1}=\frac{\textbf{I}_\frac{N_p}{N}}{\rho }-\frac{%
	\textbf{H}_{ij}^{\ast }}{\rho ^{2}}\left( \textbf{I}_{\frac{N_m}{M}}+\frac{\textbf{H}_{ij}\textbf{H}_{ij}^{\ast }}{\rho }%
\right) ^{-1}\textbf{H}_{ij}.
\end{equation}

The primal and dual residuals, which are vectors of $M\cdot N$ components that measure the convergence of the algorithm, are computed, respectively, as follows:
\begin{subequations}
	\begin{equation}
	\begin{split}
		\textbf{r}_{\textbf{p}}^{(k)}=\left(\textbf{u}_1^{1,(k)}-\textbf{v}_1^{(k)},\right.&\left.\dots,\textbf{u}_1^{M,(k)}-\textbf{v}_1^{(k)},\right.\\
		 &\vdots\\
		\textbf{u}_N^{1,(k)}-\textbf{v}_N^{(k)},&\left. \dots,\textbf{u}_N^{M,(k)}-\textbf{v}_N^{(k)} \right),
	\end{split}
	\end{equation}
	\begin{equation}
	\begin{split}
		\textbf{r}_{\textbf{d}}^{(k)}=-\rho\left(\textbf{v}_1^{(k)}-\textbf{v}_1^{(k-1)},\right.&\left.\dots,\textbf{v}_1^{(k)}-\textbf{v}_1^{(k-1)}\right.\\
		&\vdots\\
		\textbf{v}_N^{(k)}-\textbf{v}_N^{(k-1)},&\left.\dots,\textbf{v}_N^{(k)}-\textbf{v}_N^{(k-1)}\right),
	\end{split}
	\end{equation}
\end{subequations}
and their squared norms are
\begin{subequations}
	\begin{equation}
	\Vert\textbf{r}_{\textbf{p}}^{(k)}\Vert_2^2=\sum_{i=1}^M\sum_{j=1}^N\Vert \textbf{u}_j^{i,(k)}-\textbf{v}_j^{(k)}\Vert_2^2,
	\end{equation}	
	\begin{equation}
	\Vert\textbf{r}_{\textbf{d}}^{(k)}\Vert_2^2=\rho^2M\sum_{j=1}^N\Vert \textbf{v}_j^{(k)}-\textbf{v}_j^{(k-1)}\Vert_2^2.
	\end{equation}	
\end{subequations}

Equations \eqref{by_rows_and_columns_consensus}-\eqref{inversion_lemma3} combine the particularities of both previous approaches for solving the original problem introduced in Eqn. \eqref{linear_eq_with_noise}. The matrix \textbf{H} is divided in submatrices by rows ($i$ indices) and by columns ($j$ indices). For this reason, the unknown vector \textbf{u} is divided into $N$ segments [$\textbf{u}_1; \dots ;\textbf{u}_N$] and each of them is replicated $M$ times ($\textbf{u}^1_j, \dots, \textbf{u}^M_j$, for $j=1,\dots,N$). 

\begin{figure}[htp]
	\centering
	\includegraphics[scale=.55, trim = 1mm 195mm 0mm 8mm, clip]{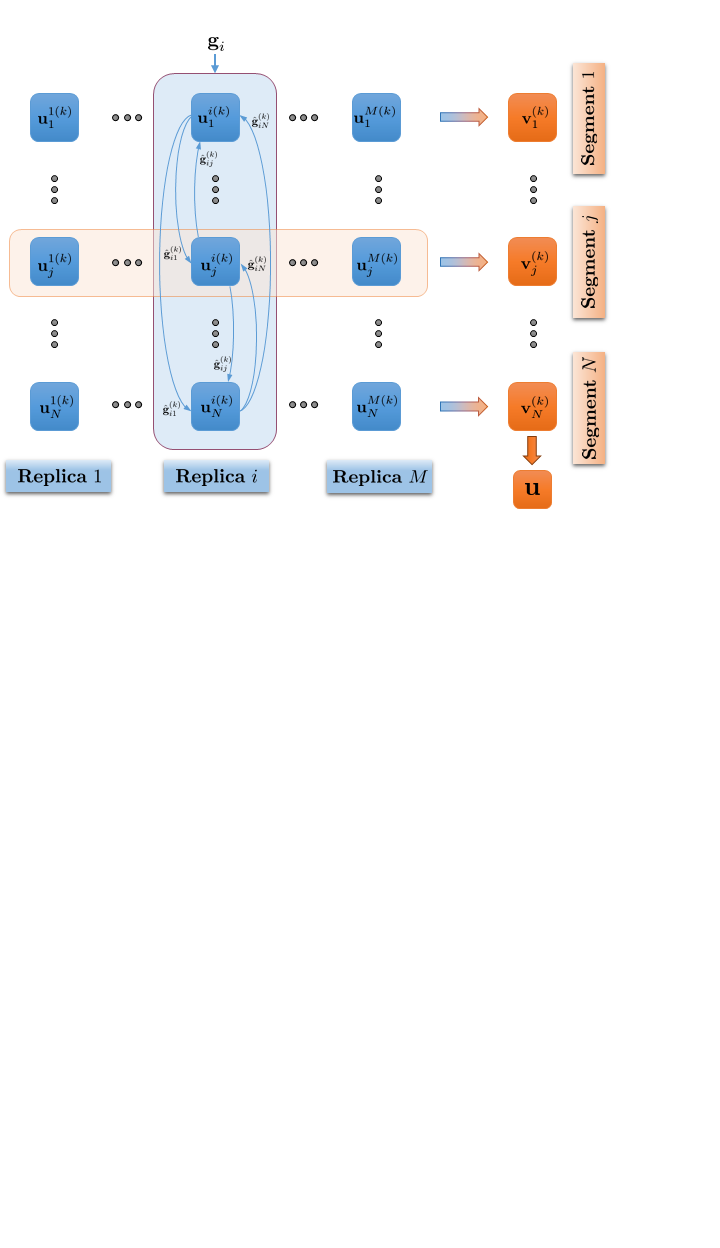}
	\caption{Schematic of the rows and columns-wise division resolution. The unknown vector \textbf{u} is divided into $N$ segments and replicated $M$ times. For a fixed replica $i$, the optimization of each sub-variable $\textbf{u}_j^{i,(k)}$ for $j=1,\dots,N$ requires the knowledge of the subvector $\textbf{g}_i$ and the \textit{estimated data} $\hat{\textbf{g}}_{iq}^{(k)}$ obtained from the previous optimizations of the remaining sub-variables $\textbf{u}_q^{i,(k)}$ for $q=1,\dots,N$ with $q\neq j$. For a given segment $j$, the sub-variable $\textbf{v}_j^{(k)}$ acts as the \textit{consensus} of all the replicas $\textbf{u}_j^{i,(k)}$, for $i=1.\dots,M$, of that segment.}
	\vspace{-5pt}
	\label{Sections_replicas}
\end{figure}

For solving this problem there are two steps in which some information need to be shared. On one hand, Eqns. \eqref{by_rows_and_columns_consensus} and \eqref{v_variable_rows_columns} show that, for a given segment $j$, $\textbf{v}_j^{(k+1)}$ acts as a \textit{consensus} variable, imposing the agreement among all $\textbf{u}_j^{i,(k+1)}$ for $i=1,\dots,M$, namely, among all replicas of the same segment. On the other hand, Eqns. \eqref{x_solution_rows_columns} and \eqref{g_rows_columns} show that, for a given replica $i$, the optimization of the variables $\textbf{u}_j^{i,(k+1)}$ for $j=1,\dots,N$, that is, the optimization of all segments of the same replica, require the knowledge of the subvector $\textbf{g}_i$, as well as the updates of the \textit{estimated data} $\hat{\textbf{g}}_{iq}^{(k)}=\textbf{H}_{iq}\textbf{u}_q^{i,(k)}\in\mathbb{C}^{N_m}$ for $q=1,\dots,N$ with $q\neq j$, from the previous iteration. This explanation is depicted in Fig. \ref{Sections_replicas}.

\begin{figure}[htp]
	\centering
	\vspace{0pt}
	\includegraphics[scale=.47, trim = 2mm 38mm 0mm 75mm, clip]{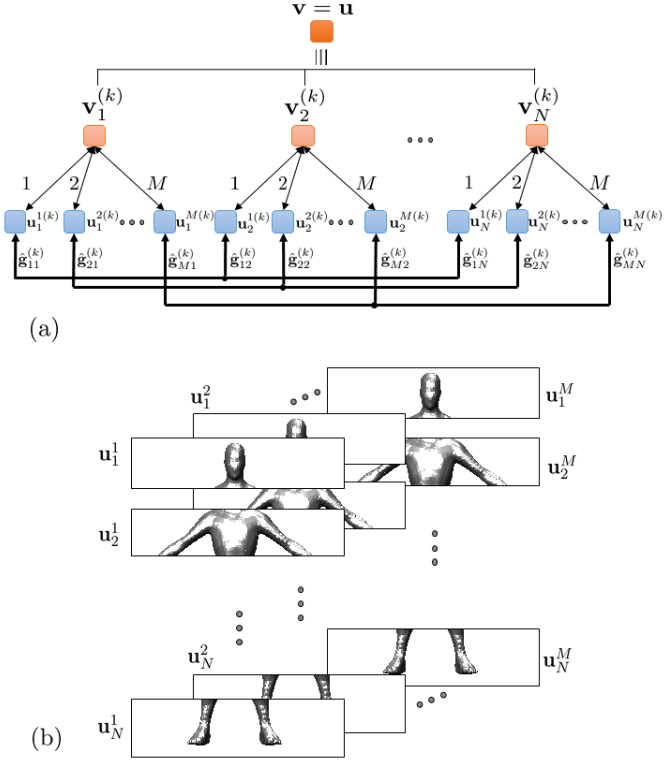}
	\caption{(a) Architecture of the consensus and sectioning-based ADMM: the problem is split into $N$ nodes, each of them acting as a central node that collects the updates of $M$ sub-nodes, computes the soft-thresholding operator of the mean of them, and then broadcast the solution again to the sub-nodes. Each sub-node shares, for each iteration, the small estimated data vector with the remaining sub-nodes that correspond with the same replica. (b) Graphical interpretation of the row and column-wise division: the image is sectioned into $N$ regions, and each of them is replicated $M$ times for performing the imaging with few data allocated to each node. The solution for each region is an average-like of all the replicas. The final imaging solution is the concatenation of all the regions.}
	\vspace{-20pt}
	\label{ByRows_Columns_bis}
\end{figure}

Therefore, as Fig. \ref{ByRows_Columns_bis}a shows, this technique can be seen as a net formed by $N$ small nodes, each of them acting as the central node for optimizing a section of the image. These nodes collects the individual solution of $M$ sub-nodes, perform the soft-thresholding averaging, and distributes again the global result to each sub-node. There are a total of $N\cdot M$ sub-nodes, each one containing a small portion of information $\textbf{H}_{ij}$ of the general matrix \textbf{H}. For a given replica $i$, all sub-nodes have to be in communication to exchange their particular \textit{estimated data} $\hat{\textbf{g}}_{ij}^{(k)}=\textbf{H}_{ij}\textbf{u}^{i,(k)}_j$. The final imaging solution is performed by connecting the $N$ different solutions from each central node, $\textbf{v}=[\textbf{v}_1;\dots; \textbf{v}_N]$.

As graphically shown in Fig. \ref{ByRows_Columns_bis}b, this technique \textit{sections} the imaging domain into $N$ small regions. For each of them, $M$ independent images are performed with less data each one. The final imaging for each region is computed as an average-like of these independent images. Finally, the global imaging solution is the re-connection of all those regions.

In this sense, this technique combines the advantages of both previous techniques: (i) by dividing by rows, the convergence process is faster since small optimizations are performed in a parallel fashion; (ii) by dividing by columns, 
small vectors have to be shared among the nodes of the same replica; and (iii) when combining the division by rows and by columns, the vectors to be exchanged among the computational nodes of the net are of a much smaller size
, reducing the communication overhead. A detailed analysis of the communication among the nodes is explained in Sect. \ref{Communications}. These two degrees of freedom enable performing the optimization in a fast and distributed fashion, making the imaging of large domains feasible.

\section{Communication among the nodes for the ADMM solution techniques}\label{Communications}
\subsection{Exchange of information for one single node}
\begin{figure}[htp]
	\centering
	\vspace{0pt}
	\includegraphics[scale=.48, trim = 2mm 224mm 0mm 6mm, clip]{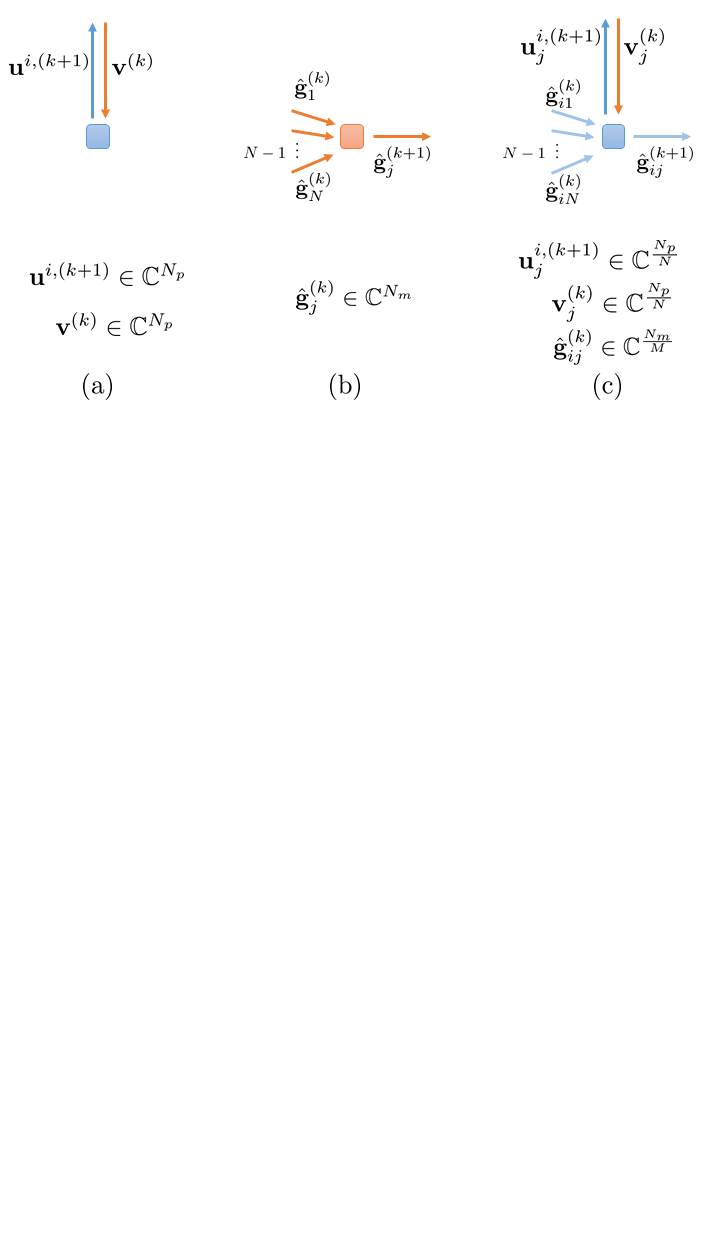}
	\caption{Schematic representation of the vectors and their lengths that are received from and transmitted by one single node at iteration $k$ when the sensing matrix of the problem is divided in submatrices (a) by rows, (b) by columns, and (c) by both rows and columns.}
	\vspace{0pt}
	\label{Communication_nodes}
\end{figure}

The three techniques studied in Sect. \ref{ADMM_formulation} present three distributed ways for finding an optimal solution of expression \eqref{original}, in which several computational nodes optimize independent sub-problems with few information allocated to each one. However, in all these three methodologies, there are concrete steps in which some information needs to be exchanged. In this section, the amount of data that is transmitted from and received by one single node at iteration $k$ is analyzed for the three techniques:
\begin{itemize}
	\item In the case of dividing the sensing matrix by rows (Fig. \ref{Communication_nodes}a), each node $i$ in the lower level has to receive the last version of $\textbf{v}^{(k)}\in\mathbb{C}^{N_p}$, and then, after the optimization, it has to send its whole new updated version $\textbf{u}^{i,(k+1)}\in\mathbb{C}^{N_p}$ to the main node (See Eqns. \eqref{x_solution}-\eqref{v_variable} and Fig. \ref{ByRows}b). The exchange of information is performed in terms of the imaging and, therefore, a total of $2N_p$ elements need to be exchanged at each iteration.
	\item In the case of dividing the sensing matrix by columns (Fig. \ref{Communication_nodes}b), each sub-node $j$ of the lower level receives the estimated data of the remaining $N-1$ nodes $\hat{\textbf{g}}_q^{(k)}\in\mathbb{C}^{N_m}$ for $q=1,\dots,N$, with $q\neq j$, and also it broadcasts its own estimated data $\hat{\textbf{g}}_j^{(k+1)}\in\mathbb{C}^{N_m}$ to the remaining nodes (See Eqns. \eqref{x_solution_columns} and \eqref{g_columns}, and Fig. \ref{ByColumns}b). Since the exchange of information is carried out in terms of the \textit{estimated data}, a total of $N\cdot N_m$ elements are shared by one node at each iteration.
	\item In the case of performing the division of the sensing matrix in both rows and columns (Fig. \ref{Communication_nodes}c), as recalled, the unknown vector \textbf{u} is divided into $N$ segments and each of them is replicated $M$ times. The sub-node $ij$, which optimizes the replica $i$ of the segment $j$ in the lower level, receives 
	the latest version of $\textbf{v}_j^{(k)}\in\mathbb{C}^{\frac{N_p}{N}}$ and 
	$N-1$ estimated data subvectors $\hat{\textbf{g}}_{iq}^{(k)}\in\mathbb{C}^{\frac{N_m}{M}}$ for $q=1,\dots,N,$ with $q\neq j$. Once the variable $\textbf{u}_j^{i,(k+1)}\in\mathbb{C}^{\frac{N_p}{N}}$ is updated, it sends it to the central node of the segment $j$, and also it broadcasts its own estimated data subvector $\hat{\textbf{g}}_{ij}^{(k+1)}\in\mathbb{C}^{\frac{N_m}{M}}$ to the remaining nodes 
	 (See Eqns. \eqref{x_solution_rows_columns}, \eqref{v_variable_rows_columns}, and \eqref{g_rows_columns}, and Fig. \ref{ByRows_Columns_bis}a). Summarizing, at each iteration, a total of $N\frac{N_m}{M}+2\frac{N_p}{N}$ elements are exchanged by one single node. In this case, the exchange of information is done as a combination of the imaging domain and the \textit{estimated data}.
\end{itemize}

Table \ref{tab:node_communication} shows the amount of elements to be received by and transmitted from one single node at iteration $k$ for the three analyzed cases.

\begin{table}[htp]
	\centering 
	\vspace{-5pt}	
	\caption{Number of elements exchanged by one single node at one iteration for the three ADMM distributed techniques}\label{tab:node_communication}
	\vspace{-7pt}	
	\begin{tabular}{|c|c|}
		\hline
		\textbf{ADMM method} & \textbf{\thead{$\#$ of elements shared \\ for one node at iteration $k$}}  \\
		\hline
		\thead{Consensus-based\\(Row-wise division)} & $2N_p$  \\
		\hline
		\thead{Sectioning-based\\(Column-wise division)} & $N\cdot N_m$ \\
		\hline	
		\thead{Consensus and Sectioning-based\\(Row and column-wise division)} & $N\frac{N_m}{M}+2\frac{N_p}{N}$ \\
		\hline	
	\end{tabular}
	\vspace{-10pt}
\end{table}

\subsection{Communication efficiency of the three distributed ADMM techniques}
In order to assess the efficiency of the communications among the nodes for the three different techniques, the amount of information received by and transmitted from one single node at iteration $k$ is compared. Since the number of pixels $N_p$ and the number of measurements $N_m$ are always known, the ratio $R=\frac{N_p}{N_m}$ is considered as the reference for the analysis of the three cases.

\subsubsection{Column-wise vs Row-wise division}
\begin{figure}[htp]
	\centering
	\vspace{0pt}
	\includegraphics[scale=.46, trim = 0mm 165mm 20mm 6mm, clip]{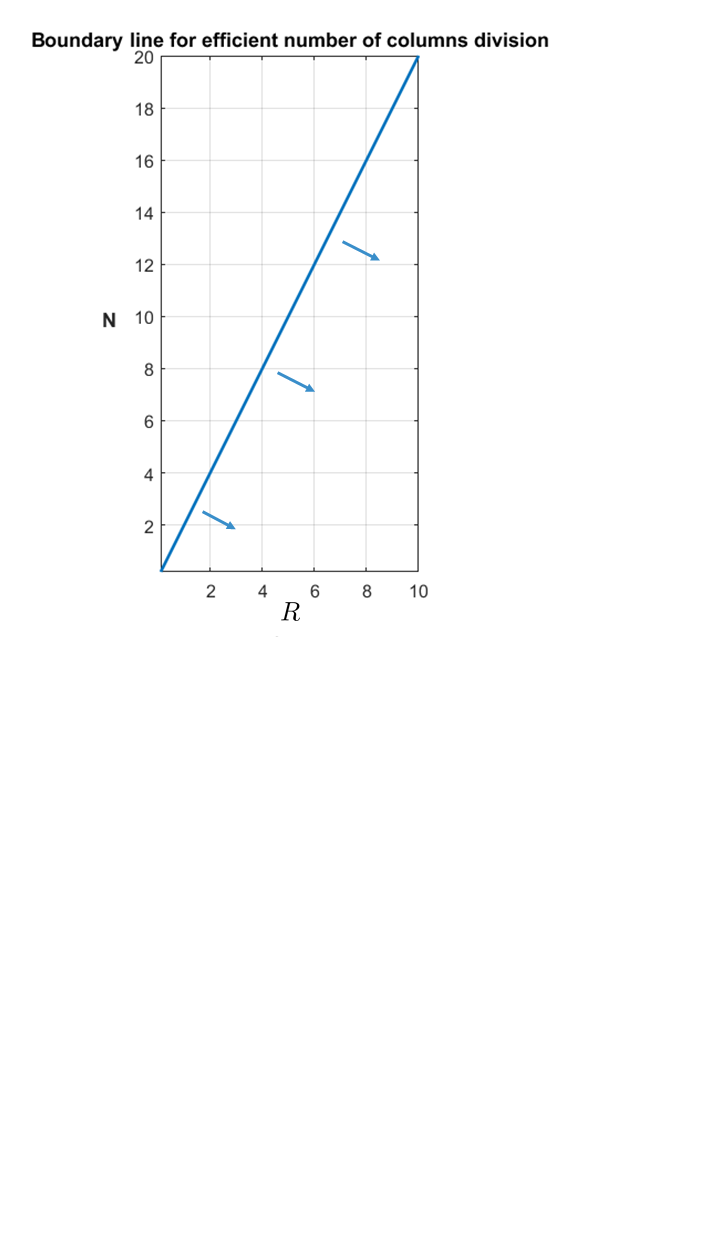}
	\caption{Boundary line comparing the efficiency of the columns-wise division versus the rows-wise division. Dividing the sensing matrix in submatrices by columns is more efficient than dividing it by rows, in terms of communications among the nodes, for the integer and positive values of $N$ that fall in the area indicated by the arrows, given $R=\frac{N_p}{N_m}$.}
	\vspace{0pt}
	\label{Efficiency_rows_columns}
\end{figure}

The columns-wise division (Sectioning-based ADMM) is more efficient than the rows-wise division (Consensus-based ADMM) in terms of communications among the nodes if the following inequation is satisfied:
\begin{equation}
	N\cdot N_m<2N_p.
\end{equation}
This implies that the number of divisions by columns of the sensing matrix has to satisfy
\begin{equation}
1<N<2R.
\end{equation}
Figure \ref{Efficiency_rows_columns} graphically represents this inequation. 

\subsubsection{Row and column-wise vs Row-wise division}
\begin{figure}[htp]
	\centering
	\vspace{0pt}
	\includegraphics[scale=.5, trim = 6mm 200mm 0mm 3mm, clip]{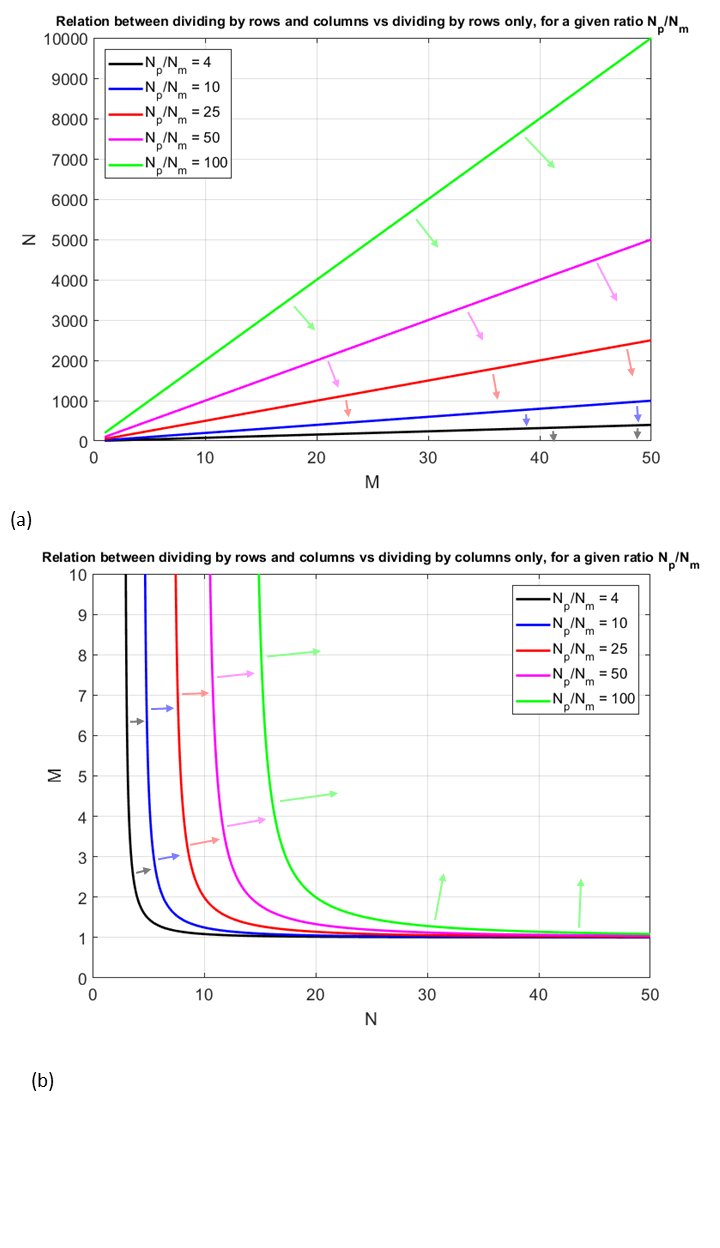}
	\caption{Boundary curves comparing the efficiency of the row and column-wise division versus the row-wise division. Dividing the sensing matrix in submatrices by rows and columns is more efficient, in terms of communications among the nodes, than dividing it by rows only, for the integer and positive values of $M$ and $N$ that fall in the area indicated by the arrows, for a given ratio $R=\frac{N_p}{N_m}$.}
	\vspace{0pt}
	\label{Efficiency_rowscolumns_rows}
\end{figure}

The row and column-wise division (Consensus and sectioning-based ADMM) is more efficient than the row-wise division (Consensus-based ADMM) in terms of communications among the nodes if the following inequation is satisfied:
\begin{equation}
	N\frac{N_m}{M}+2\frac{N_p}{N}<2N_p,
\end{equation}
which implies that
\begin{equation}
	\frac{N^2}{2M(N-1)}<R.
\end{equation}
For a given ratio $R$, the number of column divisions $N$, in terms of the number of row divisions $M$, must satisfy the following inequation:
\begin{equation}
	1<N<\sqrt{R^2M^2-2RM}+RM\sim 2RM.
\end{equation}
Figure \ref{Efficiency_rowscolumns_rows} represents this inequation for some particular ratios $R$. The division of the sensing matrix in rows and columns is more efficient than the division in rows only for those integer and positive values of $M$ and $N$ that fall in the region indicated by the arrows.

\subsubsection{Rows and columns-wise vs Columns-wise division}
\begin{figure}[htp]
	\centering
	\vspace{0pt}
	\includegraphics[scale=.5, trim = 6mm 58mm 0mm 145mm, clip]{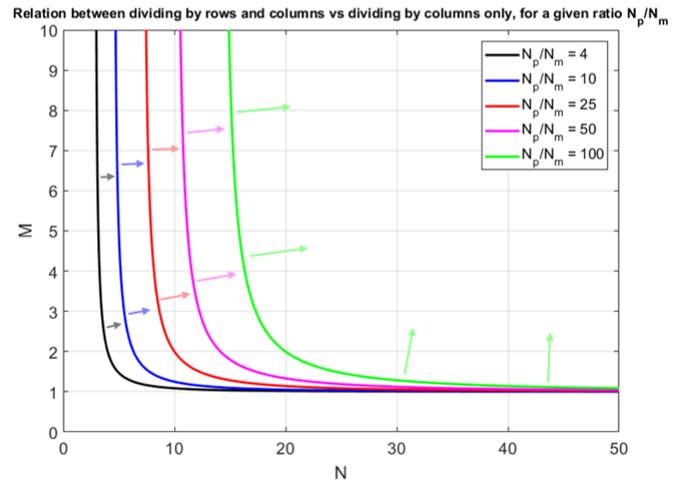}
	\caption{Boundary curves comparing the efficiency of the row and column-wise division versus the column-wise division. Dividing the sensing matrix in submatrices by rows and columns is more efficient, in terms of communications among the nodes, than dividing it by columns only, for those integer values of $N$ and $M$ that fall in the area indicated by the arrows, for a given ratio $R=\frac{N_p}{N_m}$.}
	\vspace{0pt}
	\label{Efficiency_rowscolumns_columns}
\end{figure}

The rows and columns-wise division (Consensus and sectioning-based ADMM) is more efficient than the columns-wise division (Sectioning-based ADMM) in terms of communications among the nodes if the following inequation is satisfied:
\begin{equation}
N\frac{N_m}{M}+2\frac{N_p}{N}<N\cdot N_m.
\end{equation}
This implies that
\begin{equation}
\frac{N^2(M-1)}{2M}<R.
\end{equation}
Therefore, given a ratio $R$, the number of divisions by rows $M$ in terms of the number of divisions by columns $N$ must satisfy
\begin{equation}
M>\frac{N^2}{N^2-2R}.
\end{equation}
Figure \ref{Efficiency_rowscolumns_columns} represents this inequation for some specific ratios $R$. The division of the sensing matrix in rows and columns is more efficient than the division in columns only for those integer values of $N$ and $M$ that fall in the region indicated by the arrows.

\section{Compressive Reflector Antenna}\label{CRA}
\begin{figure}[htp]
	\centering
	\vspace{0pt}
	\includegraphics[scale=.5, trim = 4mm 178mm 0mm 65mm, clip]{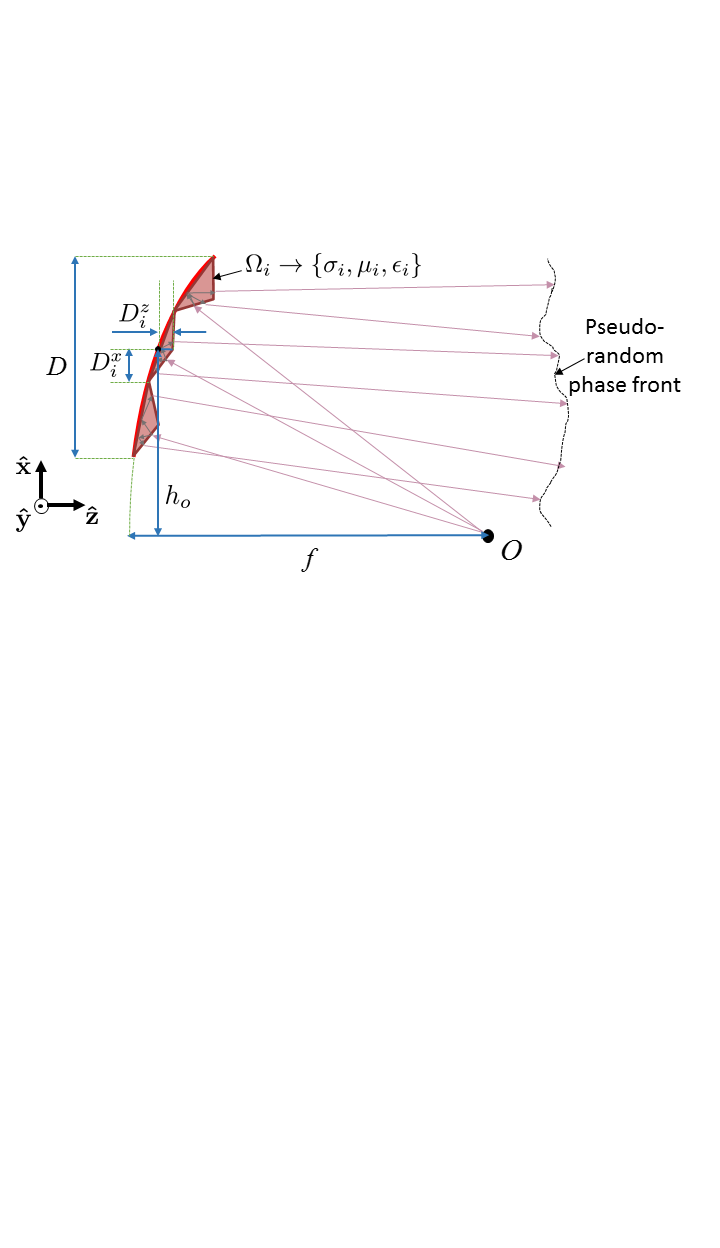}
	\caption{2D cross-section of a CRA in  offset mode. The scatterers $\Omega_i$ distort the phase front creating a pseudo-random pattern.}
	\vspace{0pt}
	\label{fig:CRA}
\end{figure}

Compressive Reflector Antenna (CRA) has been presented recently as a hardware capable of improving the sensing capacity of imaging systems in passive \cite{molaei2018interferometric,Molaei2016} and active \cite{Gomez-Sousa2016,molaei20172,molaei2016active,molaei2017high,molaei2017single} mm-wave radar applications. The CRA is built by distorting the surface of a Traditional Reflector Antenna (TRA) with some scatterers $\Omega_i$, characterized by their dimension $\{D_i^x,D_i^y,D_i^z\}$ and electromagnetic properties: permittivity $\epsilon_i$, permeability $\mu_i$, and conductivity $\sigma_i$, as it is shown in Fig. \ref{fig:CRA}. Other parameters, such as the aperture size $D$, the focal distance $f$, and the offset height $h_o$ are in common with the TRA. This distortion modifies the well-known planar phase front pattern of the TRA, creating pseudo-random patterns that can be considered as spatial and spectral codes in the near and far field of the antenna \cite{Molaei2018}. This phenomenon reduces the mutual information among the measurements, increases the sensing capacity of the system, and allows the use of CS techniques for performing the imaging of 3D objects \cite{molaei2017compressive,Martinez-Lorenzo2015,MartinezLorenzoInpress,Obermeier2016}. 

\begin{figure}[htp]
	\centering
	\subfigure[]{
		\label{fig:configuration_a}
		\includegraphics[scale=.6, trim = 28mm 190mm 0mm 20mm, clip]{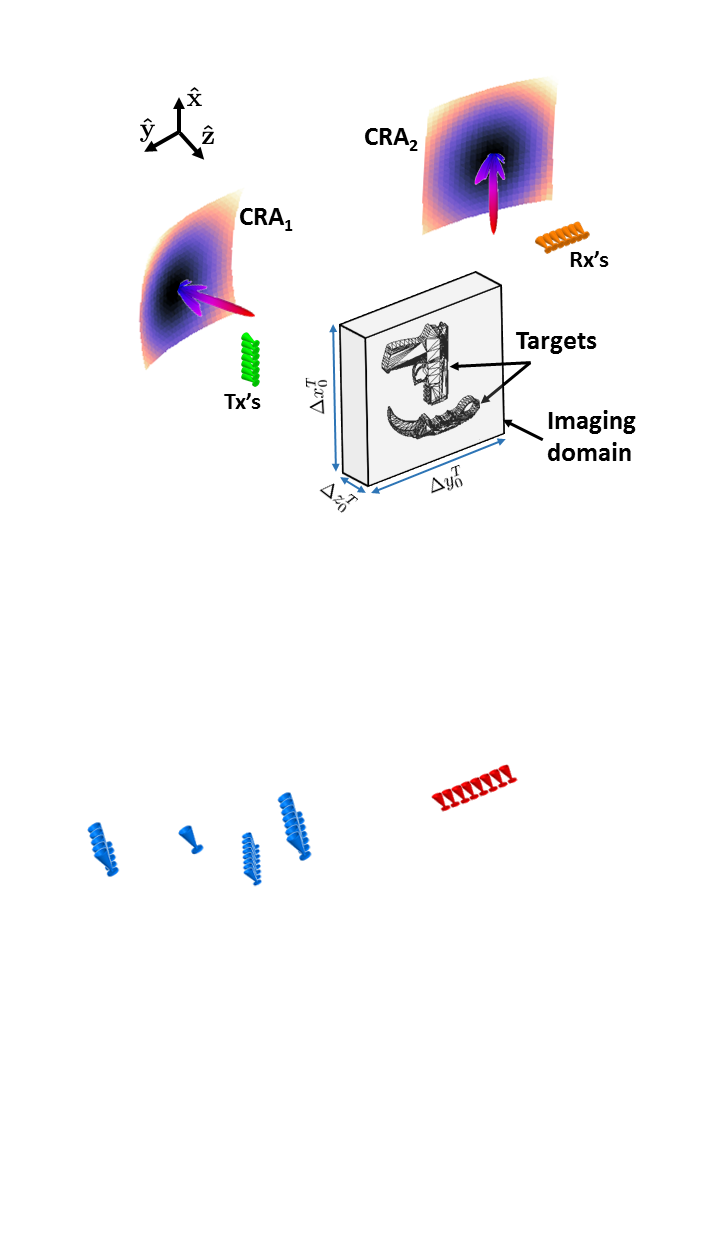}}
	\subfigure[]{
		\label{fig:configuration_b}
		\includegraphics[scale=.58, trim = 20mm 142mm 0mm -20mm, clip]{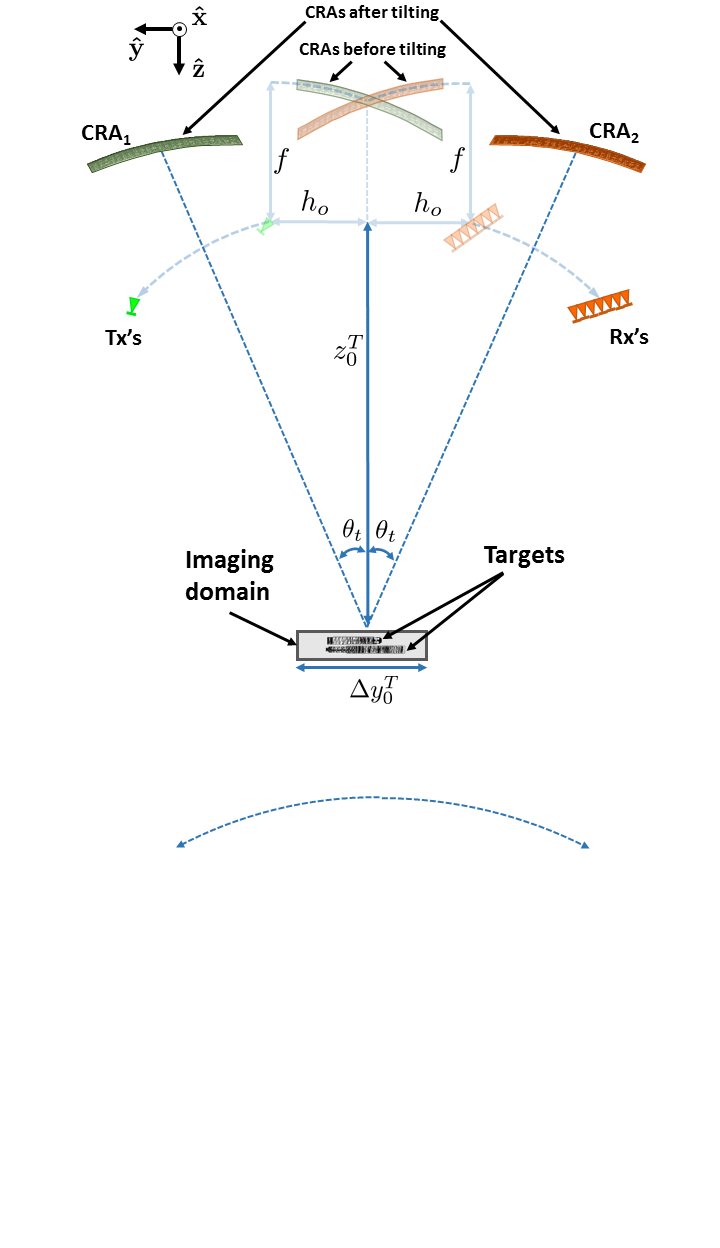}}	
	\caption{(a) Geometry of the sensing system. A linear array of transmitters feed the CRA$_1$, which illuminates the imaging domain. The field scattered by the targets is reflected by the CRA$_2$ and measured by another linear array of receivers, orthogonal to the transmitting one. (b) Top view of the sensing system. The faded CRAs and Tx and Rx arrays indicate their position before tilting. The green CRA (CRA$_1$) is tilted $\theta_t$ degrees in the $+\bf \hat y$ direction (counterclockwise), and the orange CRA (CRA$_2$) is tilted $\theta_t$ degrees in the $-\bf \hat y$ direction (clockwise).}
	\vspace{0pt}
	\label{fig:configuration}
\end{figure}

Based on the configuration depicted in Fig. \ref{fig:configuration}, $N_{Tx}$ transmitting antennas and $N_{Rx}$ receiving antennas are facing the CRA$_1$ and CRA$_2$, respectively. The signal sent from each transmitter is collected by each receiver after being scattered by the targets. The total number of measurements collected is $N_m=N_{Tx}\cdot N_{Rx}\cdot N_f$, where $N_f$ is the total number of equally-spaced frequencies used within a bandwidth of $BW$ around the central frequency $f_c$. The imaging domain is discretized into $N_p$ pixels. A linear relationship can be established between the vector of measurements $\textbf{g} \in {\mathbb{C}}^{N_m}$ and the unknown vector of reflectivity $\textbf{u} \in
{\mathbb{C}}^{N_p}$ as follows:
\begin{equation}
	\bf{g=Hu+w},
	\label{sensing_eq}
\end{equation}
where $\textbf{H}\in {\mathbb{C}}^{N_m\times N_p}$ is the sensing matrix computed as described in \cite{Meana2010} and $\textbf{w} \in {\mathbb{C}}^{N_m}$ is the noise collected for each measurement.

\section{Numerical Results}\label{Results}
The effectiveness of the three ADMM techniques is assessed by the use of CRAs for mm-wave imaging applications. Figure \ref{fig:configuration_a} shows a schematic of the configuration for the imaging problem. Two $h_o$-offset CRAs are tilted $\theta_t$ and $-\theta_t$ degrees, as shown in Fig. \ref{fig:configuration_b}. The transmitting array, arranged along the $\hat{x}$-axis and centered in the focal point of the CRA$_1$, is facing CRA$_1$; meanwhile the receiving array, linearly arranged in the YZ-plane and centered in the focal point of the CRA$_2$, is facing the CRA$_2$. The surfaces of the two CRAs are discretized into triangular patches, as it is described in \cite{Meana2010}. A scatterer is constructed over each patch, with averaged sizes of $\left\langle D^x\right\rangle$ and $\left\langle D^y\right\rangle$ in the $\bf \hat x$ and $\bf \hat y$ dimensions, respectively. The size in the $\bf \hat z$ dimension $D^z_i$ is defined as the product $\left\langle D^x\right\rangle\cdot\tan (\alpha_{t_i})$, with $\alpha_{t_i}$ being the tilt angle for each scatterer, selected from a uniform random variable in the interval $[0,\alpha_{tmax}]$, allowing a maximum tilt angle of $\alpha_{tmax}$. The material of each scatterer is considered as a perfect electric conductor (PEC), therefore $\sigma_i=\infty$. The imaging domain, where the targets are contained, is located $z_0^T$ meters away from the focal plane of the CRAs before tilting. It covers a parallelepiped-shaped volume defined by the $\Delta x^T_0 $, $\Delta y^T_0$, and $\Delta z^T_0$ dimensions, and it is discretized into $N_p$ pixels of dimensions $l_x$, $l_y$, and $l_z$. The values for all these parameters are shown in Table \ref{tab:config}.

\begin{table}[htp]
	\centering \caption{Parameters of the numerical simulation.}\label{tab:config}
	\setlength{\extrarowheight}{1.5pt}
	\begin{tabular}{|l|l||l|l|}
		\hline
		\multicolumn{1}{|c|}{\textbf{PARAM.}} & \multicolumn{1}{c||}{\textbf{VALUE}} & \multicolumn{1}{c|}{\textbf{PARAM.}} & \multicolumn{1}{c|}{\textbf{VALUE}} \\
		\hline
		\multicolumn{1}{|c|}{$f _{c}$} & \multicolumn{1}{c||}{$73.5\;GHz$} & \multicolumn{1}{c|}{$\Delta x_{0}^{T}$} & \multicolumn{1}{c|}{$30cm$} \\
		\hline
		\multicolumn{1}{|c|}{$BW$} & \multicolumn{1}{c||}{$7\;GHz$} & \multicolumn{1}{c|}{$\Delta y_{0}^{T}$} & \multicolumn{1}{c|}{$30cm$} \\
		\hline
		\multicolumn{1}{|c|}{$\lambda _{c}$} & \multicolumn{1}{c||}{$4.1\cdot 10^{-3}m$} & \multicolumn{1}{c|}{$\Delta z_{0}^{T}$} & \multicolumn{1}{c|}{$6cm$}  \\
		\hline
		\multicolumn{1}{|c|}{$D$} & \multicolumn{1}{c||}{$50 cm$} & \multicolumn{1}{c|}{$l_x$} & \multicolumn{1}{c|}{$\lambda_c$}\\
		\hline
		\multicolumn{1}{|c|}{$f$} & \multicolumn{1}{c||}{$50cm$} & \multicolumn{1}{c|}{$l_y$} & \multicolumn{1}{c|}{$\lambda_{c}$}\\
		\hline
		\multicolumn{1}{|c|}{$h_{o}$} & \multicolumn{1}{c||}{$35cm$} &
		\multicolumn{1}{c|}{$l_z$} & \multicolumn{1}{c|}{$5\lambda_{c}$}\\
		\hline
		\multicolumn{1}{|c|}{$\theta_t$} & \multicolumn{1}{c||}{$30^{\circ}$} &
		\multicolumn{1}{c|}{$N_{Tx}$} & \multicolumn{1}{c|}{$12$}\\
		\hline
		\multicolumn{1}{|c|}{$\left\langle D^{x}\right\rangle $} & \multicolumn{1}{c||}{$5\lambda _{c}$} & \multicolumn{1}{c|}{$N_{Rx}$} & \multicolumn{1}{c|}{$12$}  \\
		\hline
		\multicolumn{1}{|c|}{$\left\langle D^{y}\right\rangle $} & \multicolumn{1}{c||}{$5\lambda _{c}$} & \multicolumn{1}{c|}{$N_f$} & \multicolumn{1}{c|}{$15$}  \\
		\hline
		\multicolumn{1}{|c|}{$\alpha_{tmax}$} & \multicolumn{1}{c||}{$3^{\circ}$} & \multicolumn{1}{c|}{$N_m$} & \multicolumn{1}{c|}{$2160$} \\
		\hline
		\multicolumn{1}{|c|}{$z_{0}^{T}$} & \multicolumn{1}{c||}{$86cm$} &\multicolumn{1}{c|}{$N_p$} & \multicolumn{1}{c|}{$22500$} \\
		\hline		
	\end{tabular}
\end{table}

\begin{figure}
	\centering
	\subfigure[]{		
		\includegraphics[scale=.4, trim = 0mm 157mm 0mm 0mm, clip]{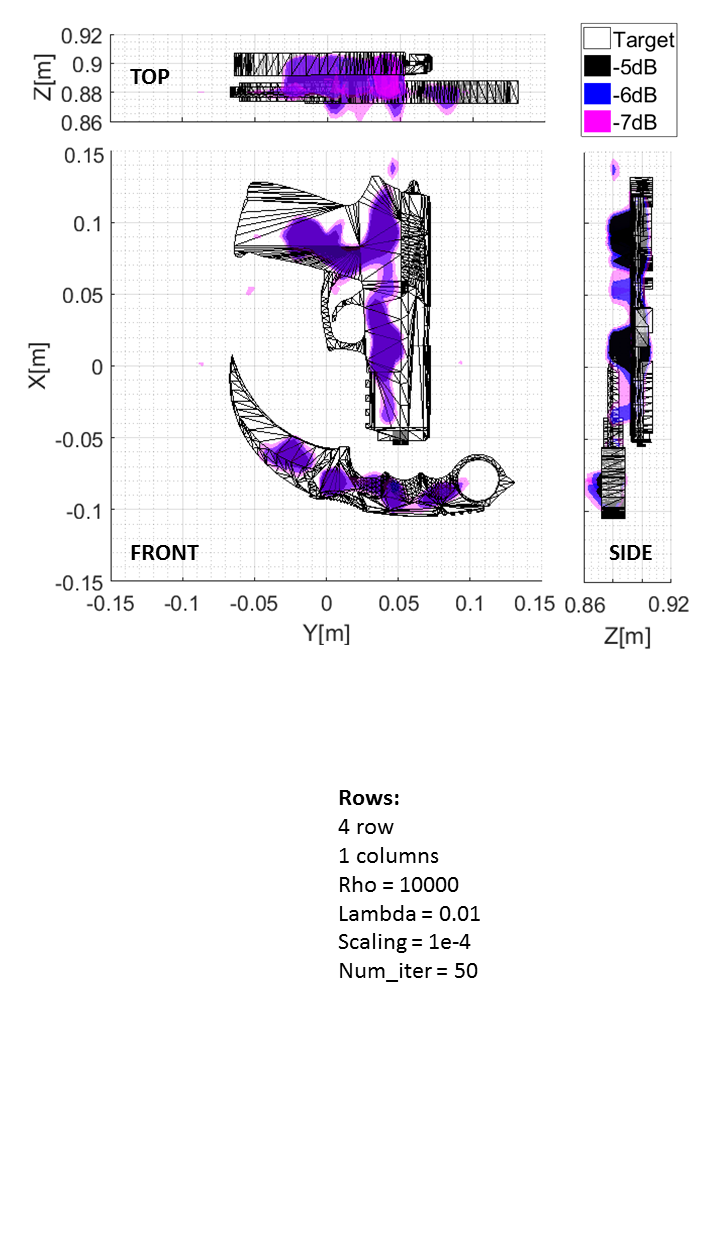}}
	\subfigure[]{
		\includegraphics[scale=.4, trim = 0mm 157mm 0mm 0mm, clip]{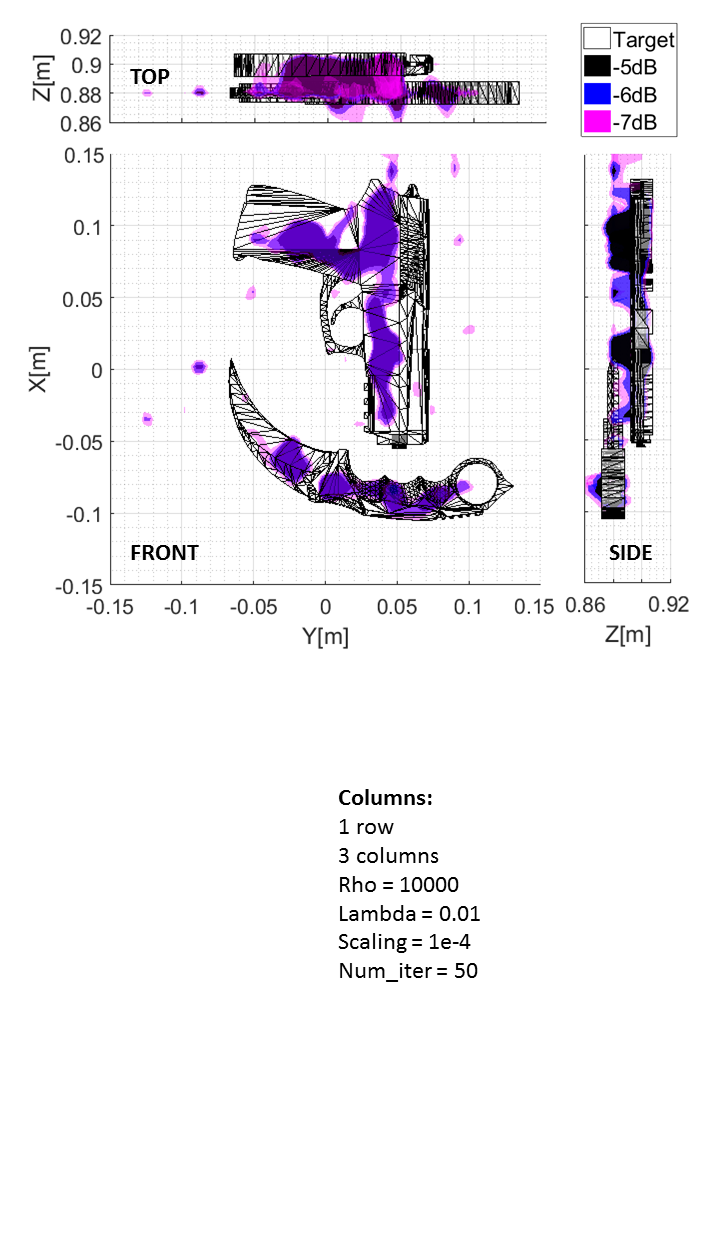}}
	\subfigure[]{
		\includegraphics[scale=.4, trim = 0mm 157mm 0mm 0mm, clip]{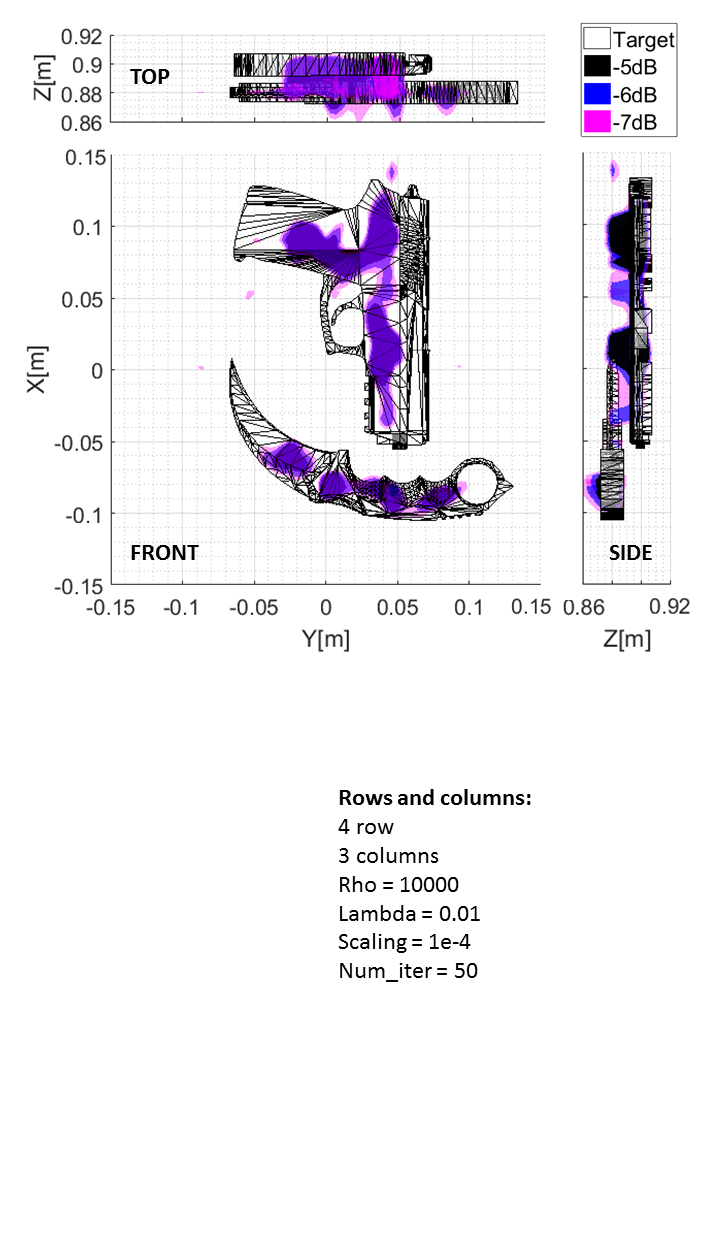}}
	\caption{Imaging reconstruction (top, front, and side views) using (a) consensus-based ADMM, (b) Sectioning-based ADMM, (c) Consensus and Sectioning-based ADMM. The targets are represented with transparent black triangles and the reconstructed reflectivity is presented in the colored map.}
	\label{fig:Imaging}
\end{figure}

\begin{figure}[htp]
	\centering
	\vspace{0pt}
	\includegraphics[scale=.46, trim = 2mm 245mm 0mm 0mm, clip]{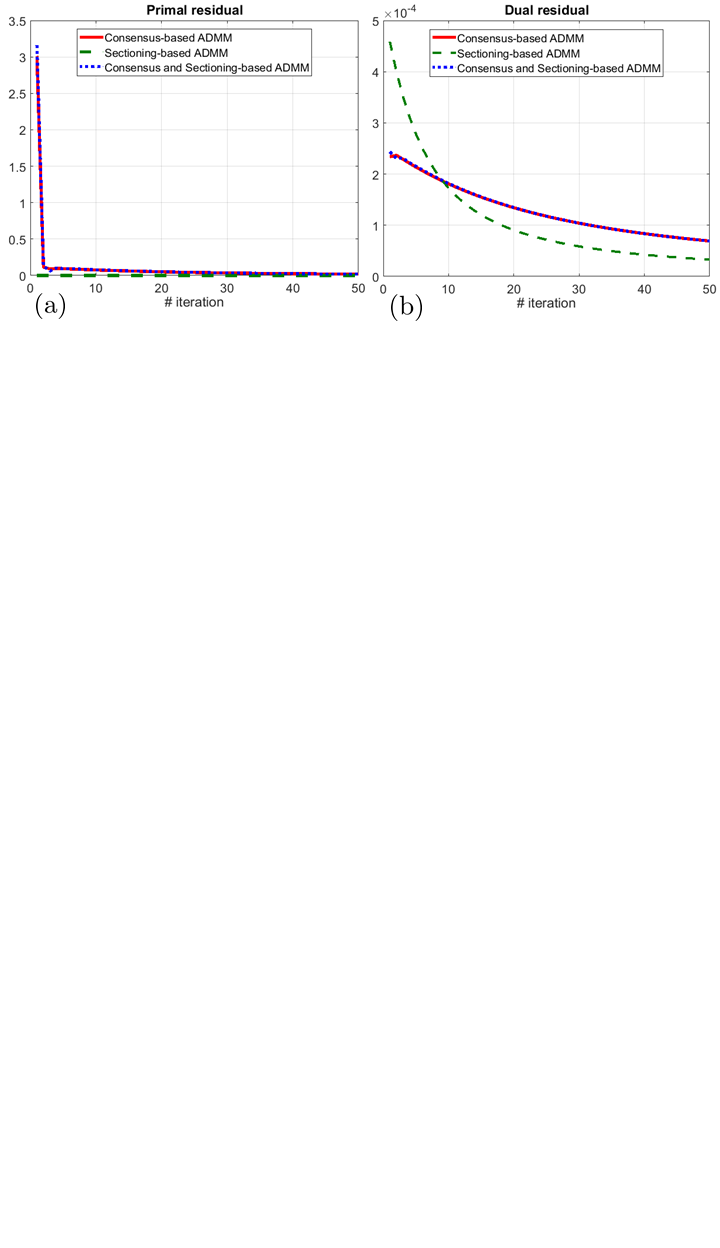}
	\caption{(a) Primal residual and (b) dual residual of the three ADMM techniques for the imaging example of Fig. \ref{fig:Imaging}. The primal residual for the Sectioning-based ADMM is almost zero since there is no \textit{consensus} for this technique.}
	\vspace{-12pt}
	\label{fig:Convergence}
\end{figure}

Figure \ref{fig:Imaging} depicts the imaging results when applying the three ADMM techniques for the following parameters: $\rho=10^5$, $\lambda=10^{-2}$, scaling factor $scl=10^{-4}$ (see Ref. \cite{Heredia-JuesasUnderreview}), and $50$ iterations. The targets correspond to a metallic gun and dagger structures located in different planes. For the consensus-based ADMM, the sensing matrix $\textbf{H}\in\mathbb{C}^{N_m\times N_p}$ is divided into $M=4$ submatrices by rows; for the sectioning-based ADMM, \textbf{H} is divided into $N=3$ submatrices by columns; and for the consensus and sectioning-based ADMM, \textbf{H} is divided into $M\cdot N=4\cdot 3 = 12$ submatrices by rows and columns. Table \ref{tab:timing} shows the sizes of the submatrices for each case, the inversion time applying the \textit{matrix inversion lemma} for those submatrices, the iterative convergence lapse time, and the total imaging time. The primal and dual convergences for each case are shown in Fig. \ref{fig:Convergence}. 

The times are computed by running an M code in a MATLAB 2017b Parallel Computer Toolbox (PCT); with a GPU Titan V, 5120 CUDA cores (1335 MHz), NVIDIA driver v390.25; in a Ubuntu Linux 16.04.4, kernel 4.13.0-36, operative system.
It can be considered that the three techniques perform the imaging in real time, especially the consensus-based ADMM, since it finds a solution in less than $1$s.

\begin{table*}[htp]
	\centering \caption{Submatrices sizes and time for the three ADMM techniques}\label{tab:timing}
	\setlength{\extrarowheight}{1.5pt}
	\begin{tabular}{|c|c|c|c|c|c|c|c|}
		\hline
		 & $\bm{M}$ & $\bm{N}$ & \textbf{Submatrices sizes} & \textbf{\thead{Submatrices sizes \\ for inverting with the \\ \textit{matrix inversion lemma}}} & \textbf{Inversion time} & \textbf{\thead{Convergence time \\ (50 iterations)}} & \textbf{Imaging time} \\
		\hline
		\textbf{\thead{Consensus-based \\ ADMM}} & $4$ & $1$ & $540\times 22500$  & $540\times 540$ & $196$ ms & $616$ ms & $0.812$ s\\
		\hline
		\textbf{\thead{Sectioning-based \\ ADMM}} & $1$ & $3$ & $2160\times 7500$ & $2160\times 2160$ & $381$ ms & $639$ ms & $1.020$ s\\
		\hline
		\textbf{\thead{Consensus and Sectioning \\ based ADMM}} & $4$ & $3$ & $540 \times 7500$ & $540\times 540$ & $248$ ms & $1171$ ms & $1.419$ s\\
		\hline		
	\end{tabular}
	\vspace{-10pt}
\end{table*}

In terms of communication among the computational nodes, Table \ref{tab:communications} shows the total amount of information that one single node has to exchange at one iteration, for the parameters of this example. It also shows the percentage of shared information reduction for the three techniques, taking the consensus-based ADMM as a reference. It is clear that the column-wise division (sectioning-based ADMM) is the most efficient technique in terms of communication, and the row-wise division (consensus-based ADMM) is the least efficient.

\vspace{-5pt}
\begin{table}[htp]
	\centering \caption{Amount of information exchanged per node at one iteration}\label{tab:communications}
	\setlength{\extrarowheight}{1.5pt}
	\begin{tabular}{|c|c|c|}
		\hline
		\textbf{ADMM method} & \textbf{\thead{$\# $of elements \\to be shared}} & \textbf{\thead{$\%$ of element reduction \\ with respect to the \\consensus-based ADMM}} \\
		\hline
		\thead{Consensus-based} & $45,000$ & $0\%$\\
		\hline
		\thead{Sectioning-based} & $6,480$ & $85.6\%$\\
		\hline
		\thead{Consensus and \\ Sectioning-based} & $12,870$ & $71.4\%$\\
		\hline		
	\end{tabular}
	\vspace{-14pt}
\end{table}

\subsection{Discussion}
Comparing the results in terms of imaging quality, imaging time, convergence, and amount of information shared among the computational nodes for the exposed example, none of the three ADMM techniques can be considered the best for all these features. The selection of one or other would depend on the feature of interest or on the physical restriction of the problem. In terms of imaging quality, even though the three techniques perform good imaging, the best option is either consensus- or sectioning and consensus-based ADMM, since they have slightly better performance. In terms of time, consensus-based ADMM has the fastest imaging time; but it is the worst when considering the amount of information exchanged among the nodes. Finally, in terms of convergence and communication efficiency, the sectioning-based ADMM is the winner; however, this method gets slower as the number of divisions gets larger, and the amount of information exchanged increases linearly. Therefore, depending on the particular needs of the problem---accuracy of the imaging, speed, computational nodes architecture, etc.---the selection of one or another method can be considered. As a general consideration, the sectioning and consensus-based ADMM technique
is always a good option, since it has more degrees of freedom that allow to get close to the best performance for the most of the features.

\vspace{-1pt}
\section{Conclusion}\label{conclusion}
Three ADMM-based techniques have been introduced to find a sparse solution of a linear matrix equation in a distributed fashion. These techniques are particularly adapted to a mm-wave imaging application. In the \textit{consensus}-based ADMM, the sensing matrix is divided in submatrices by rows, creating several replicas of the unknown imaging vector and solving them in parallel, reaching a consensus among different solutions and highly accelerating the imaging process. In the \textit{sectioning}-based ADMM, the sensing matrix is divided in submatrices by columns, sectioning the imaging in small regions and optimizing them separately, highly reducing the amount of information to be shared by one node at each iteration. Finally, in the \textit{consensus and sectioning}-based ADMM, the sensing matrix is divided in both rows and columns, segmenting the imaging and creating replicas of each region, combining the advantages of imaging quality and reduced information exchanged among the computational nodes. 

A mm-wave imaging example through the use of two CRAs has been presented. The imaging quality, the imaging time, the convergence, and the communication among the computational nodes have been analyzed and compared. The distributed capabilities of the three proposed techniques have demonstrated their ability of performing real-time imaging of metallic targets with a reduced number of measurements.

Imaging structures that could reduce the mutual information among measurements even more could accelerate the imaging process. Also, more decentralized computational architectures can reduce further the amount of information exchanged among the nodes. Future analysis will also allow to perform non-regular divisions of the sensing matrix in both rows and columns, in which those divisions may be specified by the user depending on the particular conditions, requirements, and constraints of the problem to be solved.

\section*{ACKNOWLEDGEMENT}
This work has been funded by NSF CAREER program (Award \# 1653671) and the Department of Energy (Award \# DE-SC0017614).

\bibliography{references,ADMM,CRA}

\begin{thebibliography}{10}
\providecommand{\url}[1]{#1}
\csname url@samestyle\endcsname
\providecommand{\newblock}{\relax}
\providecommand{\bibinfo}[2]{#2}
\providecommand{\BIBentrySTDinterwordspacing}{\spaceskip=0pt\relax}
\providecommand{\BIBentryALTinterwordstretchfactor}{4}
\providecommand{\BIBentryALTinterwordspacing}{\spaceskip=\fontdimen2\font plus
\BIBentryALTinterwordstretchfactor\fontdimen3\font minus
  \fontdimen4\font\relax}
\providecommand{\BIBforeignlanguage}[2]{{%
\expandafter\ifx\csname l@#1\endcsname\relax
\typeout{** WARNING: IEEEtran.bst: No hyphenation pattern has been}%
\typeout{** loaded for the language `#1'. Using the pattern for}%
\typeout{** the default language instead.}%
\else
\language=\csname l@#1\endcsname
\fi
#2}}
\providecommand{\BIBdecl}{\relax}
\BIBdecl

\bibitem{bertsekas1989parallel}
D.~P. Bertsekas and J.~N. Tsitsiklis, \emph{Parallel and distributed
  computation: numerical methods}.\hskip 1em plus 0.5em minus 0.4em\relax
  Prentice-Hall, Inc., 1989.

\bibitem{greenspan1955methods}
D.~Greenspan, ``Methods of matrix inversion,'' \emph{The American Mathematical
  Monthly}, vol.~62, no.~5, pp. 303--318, 1955.

\bibitem{akaike1973block}
H.~Akaike, ``Block toeplitz matrix inversion,'' \emph{SIAM Journal on Applied
  Mathematics}, vol.~24, no.~2, pp. 234--241, 1973.

\bibitem{boyd2011distributed}
S.~Boyd, N.~Parikh, E.~Chu, B.~Peleato, and J.~Eckstein, ``Distributed
  optimization and statistical learning via the alternating direction method of
  multipliers,'' \emph{Foundations and Trends{\textregistered} in Machine
  Learning}, vol.~3, no.~1, pp. 1--122, July 2011.

\bibitem{heredia2017norm}
J.~Heredia-Juesas, A.~Molaei, L.~Tirado, W.~Blackwell, and J.~{\'A}.
  Mart{\'\i}nez-Lorenzo, ``Norm-1 regularized consensus-based admm for imaging
  with a compressive antenna,'' \emph{IEEE Antennas and Wireless Propagation
  Letters}, vol.~16, pp. 2362--2365, 2017.

\bibitem{forero2010consensus}
P.~A. Forero, A.~Cano, and G.~B. Giannakis, ``Consensus-based distributed
  support vector machines,'' \emph{The Journal of Machine Learning Research},
  vol.~11, pp. 1663--1707, 2010.

\bibitem{mota2012distributed}
J.~F. Mota, J.~Xavier, P.~M. Aguiar, and M.~Puschel, ``Distributed basis
  pursuit,'' \emph{Signal Processing, IEEE Transactions on}, vol.~60, no.~4,
  pp. 1942--1956, April 2012.

\bibitem{tsitsiklis1986distributed}
J.~Tsitsiklis, D.~Bertsekas, and M.~Athans, ``Distributed asynchronous
  deterministic and stochastic gradient optimization algorithms,'' \emph{IEEE
  transactions on automatic control}, vol.~31, no.~9, pp. 803--812, 1986.

\bibitem{olshevsky2009convergence}
A.~Olshevsky and J.~N. Tsitsiklis, ``Convergence speed in distributed consensus
  and averaging,'' \emph{SIAM Journal on Control and Optimization}, vol.~48,
  no.~1, pp. 33--55, 2009.

\bibitem{degroot1974reaching}
M.~H. DeGroot, ``Reaching a consensus,'' \emph{Journal of the American
  Statistical Association}, vol.~69, no. 345, pp. 118--121, 1974.

\bibitem{fang2006communication}
L.~Fang and P.~J. Antsaklis, ``On communication requirements for multi-agent
  consensus seeking,'' in \emph{Networked Embedded Sensing and Control}.\hskip
  1em plus 0.5em minus 0.4em\relax Springer, 2006, pp. 53--67.

\bibitem{jakovetic2011cooperative}
D.~Jakovetic, J.~Xavier, and J.~M. Moura, ``Cooperative convex optimization in
  networked systems: Augmented lagrangian algorithms with directed gossip
  communication,'' \emph{Signal Processing, IEEE Transactions on}, vol.~59,
  no.~8, pp. 3889--3902, August 2011.

\bibitem{mehyar2005distributed}
M.~Mehyar, D.~Spanos, J.~Pongsajapan, S.~H. Low, and R.~M. Murray,
  ``Distributed averaging on asynchronous communication networks,'' in
  \emph{Decision and Control, 2005 and 2005 European Control Conference.
  CDC-ECC'05. 44th IEEE Conference on}.\hskip 1em plus 0.5em minus 0.4em\relax
  IEEE, 2005, pp. 7446--7451.

\bibitem{mota2013d}
J.~F. Mota, J.~M. Xavier, P.~M. Aguiar, and M.~Puschel, ``D-admm: A
  communication-efficient distributed algorithm for separable optimization,''
  \emph{Signal Processing, IEEE Transactions on}, vol.~61, no.~10, pp.
  2718--2723, May 2013.

\bibitem{olfati2007consensus}
R.~Olfati-Saber, J.~A. Fax, and R.~M. Murray, ``Consensus and cooperation in
  networked multi-agent systems,'' \emph{Proceedings of the IEEE}, vol.~95,
  no.~1, pp. 215--233, 2007.

\bibitem{schizas2008consensusI}
I.~D. Schizas, A.~Ribeiro, and G.~B. Giannakis, ``Consensus in ad hoc wsns with
  noisy links—part i: Distributed estimation of deterministic signals,''
  \emph{Signal Processing, IEEE Transactions on}, vol.~56, no.~1, pp. 350--364,
  January 2008.

\bibitem{schizas2008consensusII}
I.~D. Schizas, G.~B. Giannakis, S.~I. Roumeliotis, and A.~Ribeiro, ``Consensus
  in ad hoc wsns with noisy links—part ii: Distributed estimation and smoothing
  of random signals,'' \emph{Signal Processing, IEEE Transactions on}, vol.~56,
  no.~4, pp. 1650--1666, April 2008.

\bibitem{oliveri2011bayesian}
G.~Oliveri, P.~Rocca, and A.~Massa, ``A bayesian-compressive-sampling-based
  inversion for imaging sparse scatterers,'' \emph{IEEE Transactions on
  Geoscience and Remote Sensing}, vol.~49, no.~10, pp. 3993--4006, 2011.

\bibitem{beck2009fast}
A.~Beck and M.~Teboulle, ``A fast iterative shrinkage-thresholding algorithm
  for linear inverse problems,'' \emph{SIAM Journal on Imaging Sciences},
  vol.~2, no.~1, pp. 183--202, 2009.

\bibitem{Becker2011}
S.~Becker, J.~Bobin, and E.~J. Cand{\`e}s, ``Nesta: a fast and accurate
  first-order method for sparse recovery,'' \emph{SIAM Journal on Imaging
  Sciences}, vol.~4, no.~1, pp. 1--39, 2011.

\bibitem{boyd2009convex}
S.~Boyd and L.~Vandenberghe, \emph{Convex optimization}.\hskip 1em plus 0.5em
  minus 0.4em\relax Cambridge university press, 2009.

\bibitem{HerediaJuesas2015}
J.~Heredia~Juesas, G.~Allan, A.~Molaei, L.~Tirado, W.~Blackwell, and J.~A.
  Martinez~Lorenzo, ``Consensus-based imaging using admm for a compressive
  reflector antenna,'' in \emph{Antennas and Propagation Symposium}, July 2015.

\bibitem{erseghe2011fast}
T.~Erseghe, D.~Zennaro, E.~Dall'Anese, and L.~Vangelista, ``Fast consensus by
  the alternating direction multipliers method,'' \emph{Signal Processing, IEEE
  Transactions on}, vol.~59, no.~11, pp. 5523--5537, November 2011.

\bibitem{Heredia-JuesasUnderreview}
J.~Heredia-Juesas, A.~Molaei, L.~Tirado, and J.~{\'A}. Mart{\'\i}nez-Lorenzo,
  ``Sectioning-based admm imaging for fast node communication with a
  compressive antenna,'' \emph{IEEE Antennas and Wireless Propagation Letters},
  Under review.

\bibitem{HerediaJuesas2018fast}
------, ``Fast node communication admm-based imaging algorithm with a
  compressive reflector antenna,'' in \emph{Antennas and Propagation \&
  USNC/URSI National Radio Science Meeting, 2018 IEEE International Symposium
  on}.\hskip 1em plus 0.5em minus 0.4em\relax IEEE, July 2018.

\bibitem{molaei2017compressive}
A.~Molaei, J.~H. Juesas, and J.~A.~M. Lorenzo, ``Compressive reflector antenna
  phased array,'' in \emph{Antenna Arrays and Beam-formation}.\hskip 1em plus
  0.5em minus 0.4em\relax InTech, 2017.

\bibitem{Martinez-Lorenzo2015}
J.~Martinez~Lorenzo, J.~Heredia~Juesas, and W.~Blackwell, ``A
  single-transceiver compressive reflector antenna for high-sensing-capacity
  imaging,'' \emph{IEEE Antennas and Wireless Propagation Letters}, vol.~15,
  pp. 968--971, March 2016.

\bibitem{MartinezLorenzoInpress}
------, ``Single-transceiver compressive antenna for high-capacity sensing and
  imaging applications.'' in \emph{EuCAP2015}, Lisbon, In press.

\bibitem{bertsekas2014constrained}
D.~P. Bertsekas, \emph{Constrained optimization and Lagrange multiplier
  methods}.\hskip 1em plus 0.5em minus 0.4em\relax Academic press, 2014.

\bibitem{candes2008restricted}
E.~J. Candes, ``The restricted isometry property and its implications for
  compressed sensing,'' \emph{Comptes Rendus Mathematique}, vol. 346, no. 9-10,
  pp. 589--592, 2008.

\bibitem{Obermeier2016model}
R.~Obermeier and J.~A. Martinez-Lorenzo, ``Model-based optimization of
  compressive antennas for high-sensing-capacity applications,'' \emph{IEEE
  Antennas and Wireless Propagation Letters}, 2016.

\bibitem{bredies2008linear}
K.~Bredies and D.~A. Lorenz, ``Linear convergence of iterative
  soft-thresholding,'' \emph{Journal of Fourier Analysis and Applications},
  vol.~14, no. 5-6, pp. 813--837, October 2008.

\bibitem{woodbury1950inverting}
M.~A. Woodbury, ``Inverting modified matrices,'' \emph{Memorandum report},
  vol.~42, p. 106, 1950.

\bibitem{molaei2018interferometric}
A.~Molaei, J.~H. Juesas, W.~Blackwell, and J.~A.~M. Lorenzo, ``Interferometric
  sounding using a metamaterial-based compressive reflector antenna,''
  \emph{IEEE Transactions on Antennas and Propagation}, 2018.

\bibitem{Molaei2016}
A.~Molaei, G.~Allan, J.~Heredia, W.~Blackwell, and J.~Martinez-Lorenzo,
  ``Interferometric sounding using a compressive reflector antenna,'' in
  \emph{Antennas and Propagation (EUCAP 2016)}, 2016.

\bibitem{Gomez-Sousa2016}
H.~Gomez-Sousa, O.~Rubinos-Lopez, and J.~A. Martinez-Lorenzo, ``Hematologic
  characterization and 3d imaging of red blood cells using a compressive
  nano-antenna and ml-fma modeling,'' in \emph{Antennas and Propagation
  (EUCAP2016)}, 2016.

\bibitem{molaei20172}
A.~Molaei, J.~Heredia-Juesas, and J.~Martinez-Lorenzo, ``A 2-bit and 3-bit
  metamaterial absorber-based compressive reflector antenna for high sensing
  capacity imaging,'' in \emph{Technologies for Homeland Security (HST), 2017
  IEEE International Symposium on}.\hskip 1em plus 0.5em minus 0.4em\relax
  IEEE, 2017, pp. 1--6.

\bibitem{molaei2016active}
A.~Molaei, J.~H. Juesas, G.~Allan, and J.~Martinez-Lorenzo, ``Active imaging
  using a metamaterial-based compressive reflector antenna,'' in \emph{Antennas
  and Propagation (APSURSI), 2016 IEEE International Symposium on}.\hskip 1em
  plus 0.5em minus 0.4em\relax IEEE, 2016, pp. 1933--1934.

\bibitem{molaei2017high}
A.~Molaei, G.~Ghazi, J.~Heredia-Juesas, H.~Gomez-Sousa, and
  J.~Martinez-Lorenzo, ``High capacity imaging using an array of compressive
  reflector antennas,'' in \emph{Antennas and Propagation (EUCAP), 2017 11th
  European Conference on}.\hskip 1em plus 0.5em minus 0.4em\relax IEEE, 2017,
  pp. 1731--1734.

\bibitem{molaei2017single}
A.~Molaei, J.~H. Juesas, and J.~Martinez-Lorenzo, ``Single-pixel mm-wave
  imaging using 8-bits metamaterial-based compressive reflector antenna,'' in
  \emph{Antennas and Propagation \& USNC/URSI National Radio Science Meeting,
  2017 IEEE International Symposium on}.\hskip 1em plus 0.5em minus 0.4em\relax
  IEEE, 2017, pp. 847--848.

\bibitem{Molaei2018}
A.~Molaei, K.~Graham, L.~Tirado, A.~Ghanbarzadeh, A.~Bisulco,
  J.~Heredia-Juesas, C.~Liu, J.~Von~Hotenz, and J.~Martinez-Lorenzo,
  ``Experimental results of a compressive reflector antenna producing spatial
  coding,'' in \emph{Antennas and Propagation \& USNC/URSI National Radio
  Science Meeting, 2018 IEEE International Symposium on}.\hskip 1em plus 0.5em
  minus 0.4em\relax IEEE, July 2018.

\bibitem{Obermeier2016}
R.~Obermeier and J.~A. Martinez-Lorenzo, ``Model-based optimization of
  compressive antennas for high-sensing-capacity applications,'' \emph{IEEE
  Antennas and Wireless Propagation Letters. Accepted for publication}, 2016.

\bibitem{Meana2010}
J.~Meana, J.~Martinez-Lorenzo, F.~Las-Heras, and C.~Rappaport, ``Wave
  scattering by dielectric and lossy materials using the modified equivalent
  current approximation (meca),'' \emph{Antennas and Propagation, IEEE
  Transactions on}, vol.~58, no.~11, pp. 3757--3761, Nov 2010.

\end{thebibliography}
\bibliographystyle{IEEEtran}

\end{document}